# Assessment and Decision-Making in Universities: Analytics of the Administration-Staff Compromises

Valerii Semenets [1], Vagan Terziyan [2], Svitlana Gryshko [3], Mariia Golovianko [4]

[1] *Department of Metrology and Technical Expertise, Kharkiv National University of Radio Electronics, Ukraine, valery.semenets@nure.ua*

[2] *Faculty of Information Technology, University of Jyvaskyla, Finland, vagan.terziyan@jyu.fi*

[3] *Department of Economic Cybernetics and Management of Economic Security, Kharkiv National University of Radio Electronics, Ukraine, svitlala.gryshko@nure.ua*

[4] *Department of Artificial Intelligence, Kharkiv National University of Radio Electronics, Ukraine, mariia.golovianko@nure.ua*

**Abstract.** Various processes in academic organizations include the decision points where selecting people through their assessment and ranking is performed, and the impact of wrong or right choices can be very high. How do we simultaneously ensure that these selection decisions are well balanced, fair, and unbiased by satisfying the key stakeholders' wishes? How much and what kinds of evidence must be used to make them? How can both the evidence and the procedures be made transparent and unambitious for everyone? In this paper, we suggest a set of so-called deep evidence-based analytics, which is applied on top of the collective awareness platform (portal for managing higher education processes). The deep evidence, in addition to the facts about the individual academic achievements of personnel, includes the evidence about individual rewards. However, what is more important is that such evidence also includes explicit individual value systems (formalized personal preferences in the self-assessment of both achievements and the rewards). We provide formalized procedures that can be used to drive the academic assessment and selection processes within universities based on advanced (deep) evidence and with different balances of decision power between administrations and personnel. We also show how this analytics enables computational evidence for some abstract properties of an academic organization related to its organizational culture, such as organizational democracy, justice, and work passion. We present the analytics together with examples of its actual use within Ukrainian higher education at the Trust portal.

> *"We choose, we are chosen,*
> *How often the choices do not coincide!"*
> ---------------------------------------------------
> Mikhail Tanich ("Black and White," 1972)

# 1. Introduction

Making good decisions is the key to success in everything. However, it is not always clear to us how important decisions (even our own) are made or whether they induce the best-chosen options. Evidence-based decision-making is a decision-making process that is grounded on the best available scientific and empirical evidence from the field and relevant contextual evidence. This process makes the best decisions possible using the evidence (validated data and information) available and avoids decision-making that is based on gut feeling, intuition, or instinct and instead relies on data and facts.

Also, the evidence-based decision-making process is a data-driven process. The roadmap toward a data-driven decision culture is explained in Waller (2020). Companies with a strong data-driven culture have: (a) top managers who believe that decisions must be anchored in data and defined metrics; (b) employees who are code-literate and fluent in quantitative topics; and (c) data sharing policies with universal access to the key measurements. Decision makers are required to be explicit, quantitative, and transparent in making decisions (what alternatives and tradeoffs they consider for both the decision options and the approach toward the decision). Data in such companies provide a form of evidence to back up the decision-making process when the managers are well aware of each employee's performance and vice versa (MacLane et al., 2020).

Further, effective decision-making processes must not only be evidence-based (data-driven) but must also be analytical and algorithmic. Evolution of decision analytics has reached the Analytics 4.0 level—an automated, embedded analytics—which is a crucial phase in advanced analytics evolution and, to some extent, makes the decision-making processes within enterprises fundamentally algorithmic and augmented with artificial intelligence (AI) (Davenport, 2018). This trend is associated with the emergence of new professionals within companies (data sociologists), who will focus on different sociotechnical issues that arise when technology impinges on the established work practices of individuals and groups. Further development of Analytics 4.0 may induce highly automated evidence-based decision-making procedures. Due to advances in machine learning technology in recent years, AI is becoming increasingly viable not only for augmentation but also for human labor replacement, such as the work of audit committees reported by Dheeriya & Singhvi (2021). Analytics 4.0 transformation processes coevolve with generic Industry 4.0 transformation processes. Sivathanu & Pillai (2018) envisioned essential changes in human resource management and leadership style within organizations due to Industry 4.0 transformation challenges, which require new approaches, policies, and procedures for evidence-based, transparent, analytical, and algorithmic decision-making in organizations.

Evidence-based decision-making processes could be efficiently integrated with value-based decision-making given the availability of evidence (transparency) regarding the values. Value-based decision-making usually assumes that choices are made by maximizing some objective functions (either expected value for the impact of the choice or the choice value itself) and that the decision-making process unfolds across modular subprocesses, such as perception of evidence, option evaluation, and action (or course of actions) selection (Lasarus, 1997; Suri et al., 2020). Iltis (2005) attributed value-based decision-making to organizational ethics (where the values are considered within the organizational mission framework) and argued that coherence must be evident: chosen decision options and actions reflect the mission (organization's value claims). Therefore, successful implementation of value-based decision-making should induce organizational integrity because it drives an organization's choices following its values and commitments. However, the organizational values may partly come from an organization's history, and the values are also influenced by the leaders' personal values, thereby making the decision-making process largely dependent on the current leaders. Barrett (2006) argued that values are instruments to measure the organization's consciousness. Organizations begin by mapping the values of the most senior group of managers before mapping the employees' values. The senior leadership group must be aware of the available value scope and be willing to do something about them, including committing to their own personal value change, before involving the remaining employees in value assessment. If values such as continuous learning and efficiency are used on a personal template, they are classified as individual values. The purpose of customization is to ensure that the values/behaviors that are available for selection correspond not only to the societal culture but also to the type of organizational culture.

The key to a successful organization is to have a culture based on a shared set of beliefs and values that define organizational strategy, structure, and proper way to behave within the organization, which are also supported by strategy and structure. According to iEduNote (2021), organizational culture is a system of shared assumptions, values, beliefs, and attitudes that drives the collective behavior of people in organizations. That culture defines the extent to which freedom is allowed in decision-making, how power and information flow through its hierarchy, and how committed employees are toward collective objectives. Strong culture means awareness, commitment, and justice, i.e., employees know which response the top management wants from them to any situation; while employees believe that the expected response from them should be proper, they also know that rewards will be awarded for demonstrating the organization's values in their responses. Smart policies of decision transparency and awareness regarding organizational (groups, subgroups, individuals) diversity (Herde, 2020) lead to measurable impacts such as better job satisfaction (Ock, 2020), person–organization fitness (Iyer et al., 2020), organizational commitment, and job performance (Sungu et al., 2019). Srivastava et al. (2018) discussed the adaptation of personal culture to organizational culture by addressing the concern, "How do people adapt to organizational culture, and what are the consequences for their outcomes in the organization?" Two core mechanisms have been discovered that underpin cultural fit: (1) acceptance of a focal actor by her colleagues and (2) the focal actor's attachment to her colleagues and the entire organization. Distributing decision-making rights and responsibilities is also associated with the type of organizational culture (O.B.T., 2019). In an *authoritarian culture*, all the decisions are made by the top management with no employee involvement in the decision-making process. In *participative culture*, employees actively participate in decision-making (as independent individuals or as a group). Participative decision-making can follow either a shared *dominant culture* (as a mixture of different cultures) within the organization or several autonomously existing *subcultures*. According to the *strong culture* concept, employees could be loyal and have a feeling of belongingness toward the organization, thereby respecting organizational values while making the decisions. According to the *weak culture* concept, there could be no loyalty to the company from the individual decision makers (an indicator of, e.g., employee dissatisfaction). Special case of organizational culture is related to academic organization in general and universities in particular. Universities are complex social organizations with a specific culture of academic freedom and autonomy. However, over two decades the fast-evolving challenges strongly influence the primary functions of universities. Sporn (1996) analyzed the ability of university cultures to adapt to these changes and described management approaches that mirror the specific culture of a university. Various methods for assessing culture were described, a typology for interpreting university culture was introduced, and management approaches were analyzed. For administrators and researchers, this work helps explain the implications of university culture for management processes.

The special context of decision-making is when the decision objectives, processes, and options concern people (selection, recruitment, nominations, ranking, rewarding, etc.). We believe that if the decisions concern the personnel of organizations and the decision options are evaluated as a ranking list of the target-group employees based on the evidence of their work effort and achievements, then fair evaluations should consider the evidence on motivation (rewards) applied to these people. Such value and evidence-based decision-making processes in organizations are influenced by the employees' motivation (intrinsic vs. extrinsic) and corresponding balance of the commitment (individual work effort vs. individual reward). Dysvik & Kuvaas (2013) suggested that having congruent goals and corresponding values positively influences organizational behavior due to the positive relationship between intrinsic motivation and increased work effort and decision performance. Kingman (2019) discussed the effort-reward imbalance (also known as job stress) to predict mental and physical health in academic employees working in the UK universities. The study measured the interactive effects of extrinsic efforts over time vs. independent contributions of the three-reward systems (promotion, esteem, and security). Also, the interactive effects of intrinsic effort (known as overcommitment) was explored. Discovered risk factors (for mental and physical health) need interventions to reduce extrinsic and intrinsic efforts and increase rewards in the university sector. Therefore, all decisions related to people selection, nomination, responsibilities, and duty allocation must be backed up with decisions regarding fair distribution of rewards with concern for people's health. Having smart harmonious policies, organizations may benefit from the positive side of the personnel overcommitment, which is a "work passion," and concurrently minimize the dark side of the overcommitment, which is a "burn out" (Lavigne et al., 2012). As we know from the classical theory of human motivation (Maslow, 1943), while motivation is a complex, hierarchical, and dynamic concept and concurrently personal (based on each individual's perception of his or her own missions, needs, capabilities, and justice within the organization), it is individual-value-driven, which will be shown in this study. An important component of evidence-based management (and related decision-making) is evidence-based reward management (and related decision-making aiming at organizational justice). Such management is concerned with the organizational processes and practices that recognize and value employees according to the contribution they make to achieve organizational goals. According to Armstrong et al. (2010), evidence-based reward management requires consideration of all the aspects (financial and non-financial) of a reward concept. This means that the goal is not just bribing people with more pay and better benefits, but it is more about creating a working environment of trust between managers and employees with clear evidence of both distributive and procedural justice. In an academic context (particularly when considering nomination, selection, or recruitment decisions), the fact of admission often plays the role of a reward, as discussed by Niessen et al. (2017) regarding justice perception from a student perspective.

As noticed in Nair and Wayland (2008), the evidence-based and data-driven decision-making is a fast-expanding field not only in the corporate world but also in institutions of higher education. All academic organizations access enormous volumes of data from various systems. Universities have a need to make significant decisions on the investment and performance of staff and produce academic capital on a day-to-day basis. To make efficient decisions, three key principles require consideration in evidence-based decision-making: decisions must be based on fact rather than gut feeling or perception; the clear expectation by society in general for organizations to be transparent in their decision-making process; and the need for data-driven evidence and time series analysis, which allows an organization to be competitive. The quality and accountability requirements are fundamental to all these principles, enabling the universities to succeed with their journey toward improvement. Nowadays, the universities are increasingly pressurized due to the worldwide trend of massification of higher education (students have the possibility to choose better-quality education), while educational budgets are limited because of economic crises, thereby favoring universities that can maintain high quality performance with better efficiency. Waal and Kerklaan (2013) argued that an evidence-based management approach could be applied for creating high-performing sustainable universities due to the use of well-founded (evidence-based) decisions within their management processes.

Our study has been inspired by an excellent tutorial by Meijer et al. (2020) regarding decision-making procedures for recruitment, selection, hiring, and admission in educational sector. When such decisions are made based on collected evidence, it is important not only to care about the sources and correctness of the information but also about the way we manipulate information to transfer it to the decisions. What is even more important is that the derived decisions and the grounds behind them are transparent, easily accessible, and well explained (how the procedures work, examples, limitations, etc.). The tutorial argued that these conditions are not satisfied within the current practices of personnel selection (i.e., not enough valid, well-grounded, and transparent predictors; not always consistent application of weights for the predictors across different cases; absent, delayed, or incomplete feedback, etc.); thus, the performance of corresponding decision procedures and practices is far from being perfect. We also agree that the judgments made in personnel selection will be better (fairer and potentially more efficient) if the evidence is combined based on a decision rule ("mechanical" analytically grounded judgment) rather than combining the data intuitively ("holistic" judgment), like impressions from unstructured interviews.

In our study, we connect the concepts of evidence-based and value-based decision-making with the organizational culture concept using the following definition (Terziyan, 2014) as a "bridge": "Culture of an individual or a group is a systematic manner of the deliberate use of own freedom," where:

- "Systematic manner" means that the decision choice is made based on the stable well-established system of values (vector of weights of criteria for evaluating the alternatives) and its steady evolutionary dynamics (if any), and thus can be predicted with reasonable confidence and precision;
- "Deliberate use" means that either the choice is made intentionally, with understanding of the possible effects (impact) based on previously accumulated (own and others') experience and knowledge (evidence-based decision), and not based on instincts (as in the animal world) or random (chaotic) guessing;
- "Own freedom" means the freedom to choose the preferred decision alternative from the list of possible ones in certain situations.

Therefore, the stable value system completely determines the culture of its host within the evidence-based decision-making processes, and since such a system can be easily formalized in mathematical terms, it allows one to formalize, document, provide, and upon request, visualize, analyze, compute (and so on) the culture as an object for possible use in the information systems to support the decision-making processes.

In education and other domains, all decisions are made due to some kind of evidence-based "quality" assessment of the available options. Examples of possible decisions within university processes include (but are not limited to): accreditation; licensing; ranking; monitoring; testing; student recruitment; attestation; staff recruitment; choosing objectives, methods, strategies, tactics; distribution of duties, responsibilities, rewards, power, titles, budgets, grants, scholarships, prices, etc. Every evidence-based decision-making transaction is influenced by both organizational and personal contexts. Organizational context comes from the decision environment (laws, formal duties, regulations, rules, orders, instructions, and policies), and it is usually fixed, "hardcoded" and/or kept in the protected organizational "Closed World." Such a context is comparably easy to monitor, change, or reconfigure if needed by the organizational management. For an external intruder (or a hacker), it is a difficult target to influence. Personal context comes from the minds of individuals (knowledge, beliefs, capabilities, skills, experience, interests, culture, values, emotions, etc.), and it is usually trained in (and influenced by) the "Open World," kept and managed outside the organizational control and protection. Such a context is quite difficult to capture, recognize, predict, influence, and drive the change by the organizational management. For an external intruder (or a hacker), it is a relatively easy target and the most vulnerable component for corruption, cognitive hacks, and other hybrid attacks on human minds.

Schmidt and Bansler (2016) argued that understanding computing technologies in the social context of their use is the key concern of computer-supported cooperative work and human-computer interaction research; we, however, approach our objectives differently—understanding-by-computing the key but fuzzy social concepts and interactions within computational (analytical) contexts and environments—which is closer to the social computing research domain. Shared evidence, also called "awareness," is important in collaborative coordinated work in general and in collaborative decision-making in particular (Schmidt & Randall, 2016). Therefore, there is growing popularity for special digital ecosystems (called collective awareness platforms) used to provide space for sharing evidence for collaborative decision-making and provide analytical services on top of the collected evidence (Kappas et al., 2017).

In this paper, we present our study and analytics designed and implemented within the TRUST portal (portal.dovira.eu), which is a collective awareness platform used for evidence-based individual and collaborative (selection and assessment) decision-making in Ukrainian higher education (Terziyan, 2014; Terziyan et al., 2015). Regarding Meijer et al. (2020), who suggested that assessments and selection choices in the educational domain require "mechanical" analytically grounded judgments based on a decision rule, we suggest the concept of "deep" evidence (awareness) when we enable "transparent minds," i.e., personalization of the decision rule with the individual values (explicit and shared parameters, variables or weights in the decision function) to enable the evidence to pass through the prism of each decision-maker personality during the decision process. The generic schema of a deep-evidence-based decision-making process may look as follows. Assume we have some decision problems (e.g., ranking academic personnel of some universities for selecting the best option for some nominations). Assume that we have collected evidence on the academic achievements and performance of each candidate during the agreed period and that we have collected evidence on all the rewards (granted bonuses, resources, and powers) invested already to each candidate by the university. Assume that we have a clearly defined analytical function (also known as the decision rule), which computes the score for each candidate based on the collected evidence (achievements vs. rewards). Assume that the variables of that function (e.g., weights of importance of each achievement type and each reward type) are fixed and known in advance for all applicants during the whole lifecycle of the decision process—from the call to the winner announcement. All these correspond to the concept of evidence-based "mechanical" decision-making. Further, assume that everybody in the selection committee and every applicant provides public additional evidence regarding their personal view on the decision rule parameters (weights) as a reflection of their personal value system. In this case, we would have as many decision functions for analytical manipulations as the number of persons involved. Such additional evidence (and analytics capable of capturing additional value out of it) makes the difference between just evidence-based and deep-evidence-based decision-making. Therefore, the term "deep" here means additional information on mental models (at least the "values" as weights for multicriteria individual decision-making) to enable real transparency in decision-making processes that will positively affect organizational justice and passion at work, thereby increasing organizational performance. The TRUST portal is based on the concept of Analytics 4.0 and provides space and support for deep-evidence-based decision-making. In the paper, we present some subsets of such analytics as part of the Portal services. Our study's objectives are related to the following research questions:

- How can we benefit from making explicit the personal values regarding achievements and rewards during the evidence-based assessment, ranking and selection decision processes in academic organizations, and what would be the appropriate analytics?
- Can we provide computational evidence for some abstract properties of an academic organization related to its organizational culture, such as organizational democracy, justice, and work passion, and what would be the appropriate analytics?

Further content of this paper is organized as follows: In Section 2, we argue on why the (deep) evidence on achievement, rewards, and values is important and analytically synergized when assessing and selecting academic personnel; In Section 3, we describe formalization of the deep evidence used in our study and core (atomic) analytical procedures for its processing; In Section 4, we provide the set of different complex analytical procedures to support (deep) evidence-based assessment (also self-assessment) and ranking personnel regarding different processes (driven by value systems of administration, individuals, subgroups, whole personnel, etc., with different schemas for compromises and integration of different ranking lists), exemplified by implementing and using these procedures on top of the TRUST portal; In Section 5, we emphasize the analytics and related assessment and decision-making processes, which use the evidence about rewards (reinforcements) to compute several useful features and collisions among ranking lists (as an analytics of "justice") regarding individuals and organization; In Section 6, we provide formalization and computational procedure for the concept

"work passion" as an important assessment and selection criteria, which is based on evidence regarding intrinsic vs. extrinsic motivation of people within the academic organization; and we conclude in Section 7.

## 2. Achievements vs. rewards as a balance of values

Universities have good reasons for looking for new business models (Ibrahim & Dahlan, 2016). The high cost of the traditional education programs, longer study, a long return on investment in higher education, and the rapid obsolescence of skills and the diversification of educational service providers force universities to rethink their management and look for new sustainable ways to adapt to the changes in people's within the wider economic and societal ecosystems (Jabłoński et al., 2020). Nowadays, one can observe the development of fundamentally new ways of acquiring knowledge and organizing education: networked communities (Dolle et al., 2013; Sobko et al., 2020), educational hubs and clouds (Wagh et al., 2020), flipped classroom (Burke & Fedorek, 2017), lifelong learning (Cronholm, 2021), multi-format open online resources such as MOOC (Guerrero, 2021) or SPOC (Ruiz-Palmero, 2020), Education-as-a-Service (Qasem et al., 2019), especially within the post-COVID context (Ewing, 2021), etc.

Sustainable organizational advantages in the higher education domain are now replaced by the organizational agility associated with developing adaptive and changeable settings. Related to this organizational behavior is the performance-oriented interaction of people (individuals and groups) within an organization. Such behavior depends on the organizational culture conceptualizing how people perceive and describe their work environment. Regulating attitudes and behaviors of people in an organization can always be assessed, either from the forming-process viewpoint of organizational culture or from the effectiveness-result viewpoint of joint activities.

Disruptive educational models and technologies are changing universities' goals and structures. They require new approaches to decision-making generally and particularly to evaluation and assessment that would reveal previously unknown, newly emerging options and opportunities of the new educational landscape. Recent studies (OECD, 2019) have noted that using evaluation and assessment is increasing and becoming more complex across education systems, and the priorities are shifting toward restoring trust, addressing inequalities, and using digital technologies responsibly. The viability of an organization is determined by the speed and direction of its development, which in many cases goes beyond the local level, becoming a factor in societal development (Cordoba & Midgley, 2008). Various methods of development support, differing in basic principles, elements, and techniques, demonstrate a common instrumental approach—all of them are based on evidence collection, assessment, and decision-making. An assessment culture is important for an organization to identify the potential for improvement and systemic change. Any assessment, as an analytical and transparent procedure, must meet accuracy, reliability, and validity requirements, which are especially difficult to ensure in social systems because people are one of the most complex and conflicting resources in an organization. Humans are imperfect and tend to make mistakes. People can do unexpected things and always have their own opinions. The judgments, assessments, and actions of people are subjective. Not surprisingly, such a resource is extremely difficult to assess and manage, especially in the cases when some people are assessing and selecting among other people or even among themselves.

Personnel is a unique, intelligent asset if managed properly. Acceptance by the employees of the goals, objectives, needs for change, and "rules of the game" of an organization accelerates critical business processes and enables complex organizational changes. All these cannot be achieved only by direct management, orders, and instructions, especially in academic organizations, which used to have a spirit of freedom. Serious motivation is needed to make the employees spend their energy not only on performing their business as usual functions, but also on creatively achieving the organizational development goals. Otherwise, the employees perceive the initiatives of the administration only as an additional burden and seek to minimize once own efforts. That is why the procedures of assessment and the motivation of the personnel should be seen as an integral working tool of the administration.

Motivation in the workplace is used to induce employees' improvement of those results that contribute to achieving organizational goals. The goals and objectives common to the entire team require a solidary contribution from each employee. Therefore, most motivational systems aim at fairness (that is, from the standpoint of organizational interests) to assess employees' efforts and, following this, fairly stimulate (reward) them, considering that motivation mechanisms are based on subjective abstract intangible categories that each person feels individually.

In most cases, managers striving to be fair try to avoid subjectivity in assessing employees' performance, formalizing the assessment criteria and rules for distributing remuneration (Figure 1a). A motivation system is considered fair if it

- Gives the most objective and accurate assessment of the results of the work of employees.
- Objectively and impartially distributes remuneration following the results obtained.

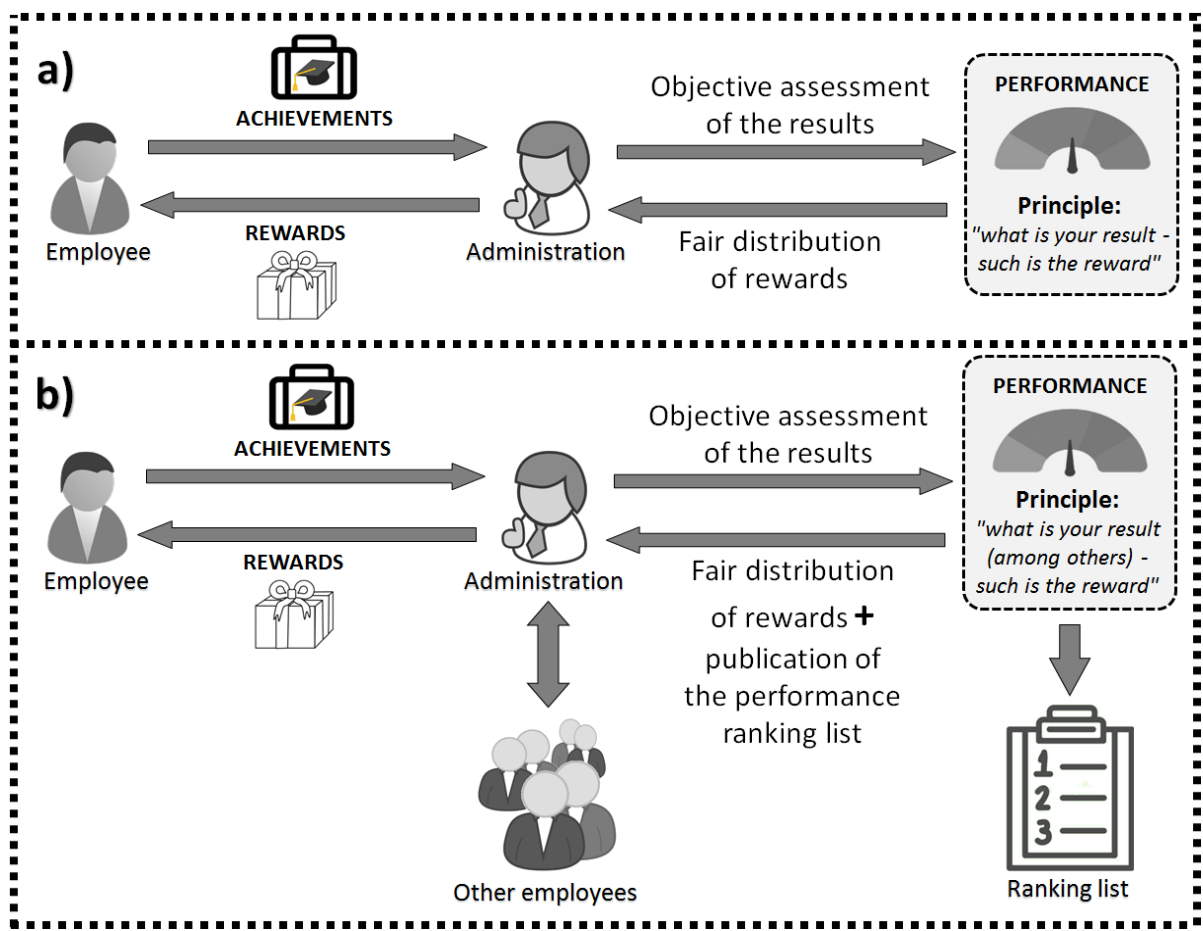

**Figure 1**: (a) The most common understanding of fairness in motivational systems; (b) Improved staff motivation scheme that enhances incentive objectivity due to transparency

More advanced systems of motivation connect the principle of transparency to this mechanism (Figure 1b). By publishing employee achievements within the ranking list with the achievements of others, the administration hopes to convey to everyone the objectivity of their decisions.

Concurrently, the following assumption seems to be obvious: low positions in the result ranking should awaken competition in the team, force employees to reconsider their attitude to work duties, and generate a desire for higher results, following the example of more successful colleagues. Therefore, the concept of "fairness" of motivation is usually associated with the concept of "objectivity" of motivation. All employees should be satisfied with this approach application and follow its logic in their activities. However, this often does not happen in practice: the motivation system has been introduced, and the staff still does not make much effort. Why does the principle of an "objective justice" fail in motivational systems? This is because objectivity contradicts both the very nature of the concept of justice and the theory of rational human behavior. Fairness on a personal level is purely subjective. Each person individually feels it, and based on these subjective feelings, assesses the situation in the coordinates "fair–not fair." You can give reasonable arguments and substantiate the objectivity of assessments, but if a person still feels injustice, all arguments, principles, and correctly built mechanisms collapse.

Now let us imagine the system of objective motivation through the eyes of the employee. For him (with respect to "her"), the organization is, first, his workplace. There a person works, that is, spends his energy and his time to "get" some useful results for the organization. Because both time and energy are a limited resource in a person's life, he does not just work: a person literally donates part of his life to the organization. Also, any employee feels this moment especially acutely. However, the employee's observation focus is much wider—he does not just work at his workplace; he works in a team. Accordingly, he sees not only his own work, but also the work of his colleagues (or is informed about it through formal and informal communications). It seems to a person that he can accurately assess how well his colleagues are doing, since he can observe this directly.

This is where distortions in subjective assessments of a person begin:

- first, it is difficult for an employee to objectively evaluate and compare his efforts and the efforts of other people: his work will always be evaluated higher since his achievements are gained "by sweat and blood," and it is impossible to feel how other people's achievements are gained, they can only be realized;
- second, it only seems to a person that he objectively sees all the work of his colleagues; he observes only some of the work from his position; also, as a rule, employees are poorly aware of all the achievements of each of their colleagues and the efforts expended on this.

A person's judgments become the basis of his rational behavior, during which he acts optimally to obtain the best result for him because this result he often considers as the benefits (remuneration) that he receives from the organization. There is also an asymmetry in how employees perceive these benefits:

- the concept of "remuneration," as a rule, is interpreted by the employee much broader than by the administration: people tend to discuss and evaluate not only salaries and bonuses, but also the distribution of work duties, positions, titles, the structure of working hours, gaining access to corporate resources, etc.;
- the value of remuneration is often assessed differently: the administration may consider the bonus a good remuneration, and the employee perceives it as an insufficient assessment of his work;
- remuneration is much more often (compared with work issues) become the object of curiosity and the subject of informal discussions in the team;
- focusing on a lack of information only feeds subjectivity in assessments.

If the rewards received are valueless to the employee himself, then the employee's rational behavior can induce "negative" optimization, motivating him to drift toward saving efforts in the workplace. All these observations induce two important conclusions:

1) The category of fairness is subjective, it reflects a person's perception of the balance (or imbalance) of his own results and costs compared with the results and costs of other employees. This conclusion also has a theoretical justification in the form of Equity Theory (Adams, 1963);

2) Justice is a relative category: people do not feel absolute justice in the form of a universal standard, but a measure of injustice; that is, they evaluate how much more justly or less justly they were treated.

Therefore, we formulate the task of assessing personnel not as building a fair system of motivation (which is impossible due to the subjective nature of justice), but as a fight against feelings of injustice. When forming the personnel motivation system, we also adhere to the premises of the Equity Theory (Adams, 1963):

- People's perception of justice (injustice) is more important than objective indicators;
- Equity is considered a balance (equivalence) of spent and received;
- For the employee, the balance between his own efforts and the received remuneration is important;
- For the employee, the balance between the ratio of his efforts and rewards and the ratio of efforts and rewards of other employees is also important;
- Injustice is an imbalance between the ratio of one's own efforts and rewards and the ratio of efforts and rewards of other colleagues;
- Access to objective information about each other (regarding performance and remuneration) can significantly reduce the distortion in the subjective perception of fairness.

We suggest moving away from the hierarchical structure of the motivation system, where the employee produces results, and the administration evaluates and rewards.

Instead, we have developed a peer-to-peer motivation system in which both the employee and the administration are equal partners:

- The employee publishes the results achieved by him during the period as an impact (that is, something that has a certain value for the organization), and the administration evaluates the proposed impacts following the administrative value system.
- However, the administration also publishes the issued rewards as an impact (that is, something that has a certain value for employees), and employees evaluate the proposed impacts following their individual value systems.

This article shares the experience of National University of Radio Electronics (NURE), one of the Ukrainian leading universities, which is an active user of the collective awareness platform (TRUST portal mentioned before) for improving assessment and selection processes. We extend the concept of awareness so that the assessment and selection analytical procedures consider not only evidence on achievements but also the evidence on the rewards and, what is more important, the analytics enables looking to the inputs (heterogeneous evidence) and inferred outcomes (decisions) through the prism of personal value systems. This experience we report as a demonstration and proof-of-use of the analytics suggested in this paper. We offer an analytical evidence-based assessment component within the TRUST portal, where the interaction "assessed–evaluator" is brought to a new level of motivation thanks to the mechanisms inherent in it:

- Trust (supported through the collective awareness platform);
- Compromise decision-making based on group decision-making methods;
- Involvement: supported by participative decision-making tools, such as representative participation (when partial responsibility for making a decision is delegated to the experts on the team) and participative leadership (when certain powers are delegated to the entire community);
- Social mobility: through the implementation of the genetic algorithm "Social lift" in the rating procedure;

- Social responsibility based on personnel self-ranking procedures, group self-assessment of the academic community;
- Pluralism of opinions—through a compromise decision-making between the evaluated and the evaluator (in the procedures weighted democracy, expert democracy, compromise dichotomy);
- "Fairness in the workplace"—as a result of the collision of ranking lists and analytics of justice procedures;
- "Work passion" which is based on policies of balancing with the intrinsic vs. extrinsic motivations of the personnel.

The analytical toolset is intended to be capable of extracting non-trivial and practically useful knowledge from a small amount of data (achievements, rewards, and value systems). The versatility of the developed assessment tool lies in the fact that it can be used in any contexts "assessed–assessor," including "student–teacher." To pilot the development, a procedure for evaluating academic personnel was chosen, during which the interaction "university management–academic personnel" was analyzed. This choice was made for several reasons:

- A fair assessment of the teacher's or researcher's work is no less important than a fair assessment of the student since the quality of educational and research services is inseparable from the subject (that is, from the teacher/researcher who provides it);
- To develop a new assessment culture in students, it must first be examined in assessors (in teachers/researchers);
- University HR management may need a new mechanism to stimulate and motivate academic staff connected with the transition from hierarchical university business models to peer-to-peer networking models.

We present the results of testing the assessment tool as a set of personnel ranking options:

1) Ranking the achievements of personnel following the administrative value system (indicator of the performance of management tasks);

2) Self-assessment of personnel both for individual and group value systems (indicator of the state of the academic environment);

3) Compromise rankings based on various combinations of opinions of the parties (a method of democratic decision-making);

4) Clash of opinions of administration and staff regarding achievements and rewards (equity diagnostic tool).

The "counter-assessment" context, which we apply in the analytics, enables making choices based on the balance "employee for the organization" vs. "organization for the employee," which allows us to formalize the concept of organizational justice. Weighing all value parameters following value systems (administrative or individual personal for each employee) ensures that the subjective factor is considered. A subjectively important achievement will "cost" more for the administration, just as the subjectively important remuneration will "cost" more for the employee. The relative nature of equity requires that not the absolute parameters of motivation be compared, but their share in the total mass (the employee's share in the total mass of the organization's results; the employee's share in the total mass of remunerations). Therefore, we believe that the assessments should be the normalized values of remuneration and results.

The academic environment has several advantages for conducting such an experiment.

- The results of the work of academic staff are clearly defined by framework and regulatory documents (this is the number of scientific articles in indexed journals, dissertation management, citations, etc.) The list of such academic results is the usual assessment criteria used in the academic environment to assess impacts at different levels (from the national higher education system to individual universities and professors);

- The results of the work of academic personnel are easily verified thanks to scientific portals and research networks, official archives of universities, resources, etc.;

- The remuneration of academic workers is also quite obvious: ranging from salary and benefits, career growth, positions to various honors and titles;
- There is a direct correlation between academic achievements and the capabilities of the university (resources on which the value and types of rewards depend)—this is provided by the principle of academic independence and affects the attraction of students, grant funding of university projects, mechanisms, etc.

Figure 2 shows the generic schema of our approach. Here, we see that all the evidence (regarding the personnel achievements and the rewards with the organization) is considered and modified regarding personal values, and such personalized (deep) evidence is processed by the transparent assessment analytics toward making decisions, ranking people, and computing important indicators for the organizational management (performance, justice, work passion, etc.).

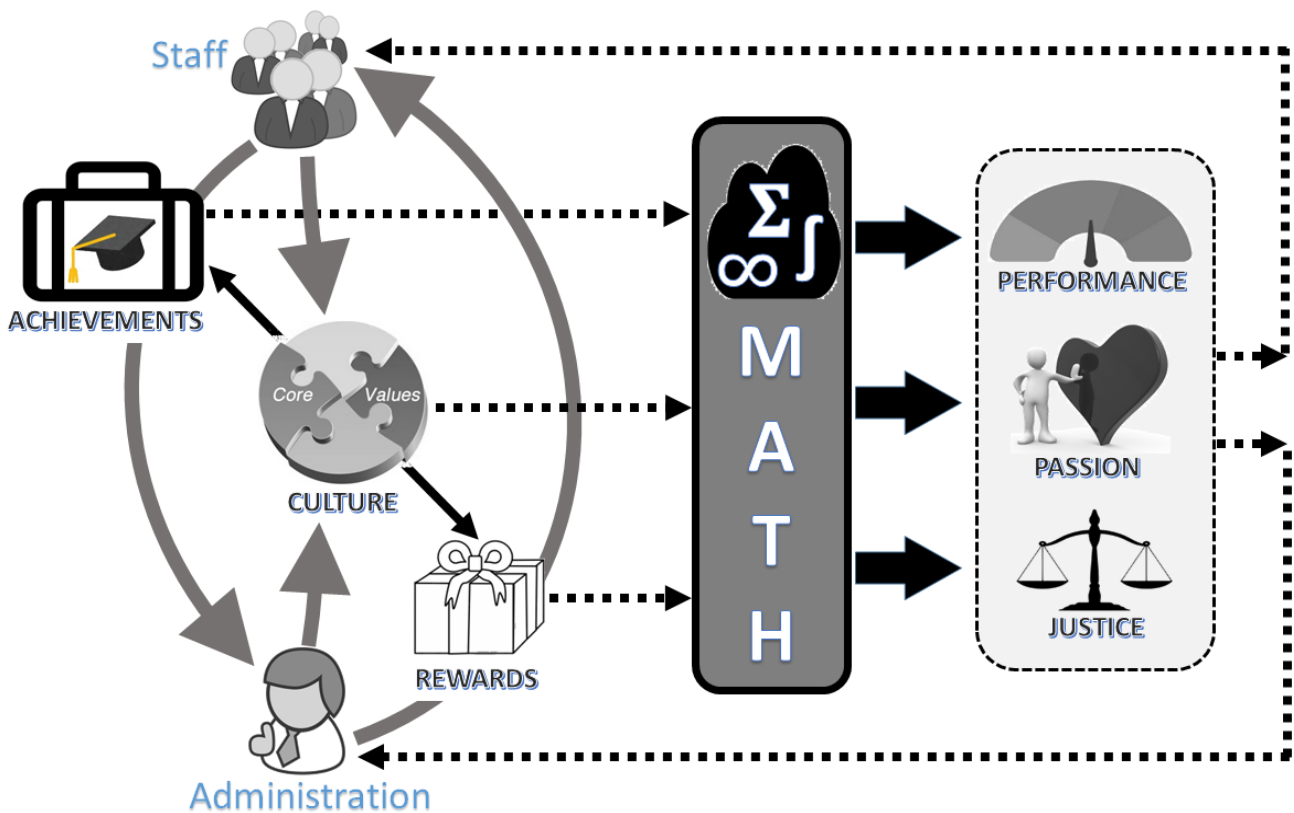

**Figure 2**: Generic schema of our approach toward deep-evidence-based analytics as a component of management structure of a university

# 3. Inputs and Key Atomic Procedures

Assume that we have some (academic) organization, which personnel comprises $m$ (academic) staff members (object of assessment). In our further experiments, we used data from Kharkiv National University of Radio Electronics (NURE).

Assume that each staff member is being assessed based on his/her achievements during a certain period. All the accountable achievements are divided into $n$ predefined categories. The contribution of each staff member is the number of documented (registered) achievements of each $n$ category. The set of categories is agreed in advance and may include, for example, the number of articles in refereed journals, the number of supervised and graduated masters, increase in the Hirsch-index of academic publications, etc.

All the contributions' statistics from all the staff members of the organization are represented by matrix $Rp$ (<u>R</u>esults, <u>p</u>ersonnel). The matrix size is $m \times n$, i.e., $Rp(i,j) = 5$ would mean that $i$-th staff member has registered 5 his (her) achievements of category $j$.

We will present various analytical procedures, which start from the $Rp$ matrix and capture various additional hidden properties of the organizational statistics. We tested the procedures on real organizational data with $m \approx 1000$ and $n \approx 50$. For the compactness and visibility of the further figures (where we will demonstrate the work of the procedures), we cut of part of that data up to $m = 30$ and $n = 4$, i.e., we selected representative set of 30 staff members with anonymized identifiers (the letters used for the abbreviated names has nothing to do with the real names) and 4 popular categories of the reported academic achievements. These selected categories are: the Hirsch-index (integer number) of academic publications from Scopus (HI); the number of supervised Candidate-of-Sciences dissertations defended during the reporting period (Head KD); the number of supervised Doctor-of-Sciences dissertations defended during the reporting period (Head DD); the number of refereed academic publication (e.g., value 1 here means one top level publication without coauthors and corresponding fraction of 1 is applied for the coauthored publications) (Papers). Figure 3 shows the [$m \times n$], i.e. [$30 \times 4$] matrix $Rp$ used for further examples.

| # | staff | HI | Head DD | Head KD | Papers |
|---|---|---|---|---|---|
| | Matrix Rp [30 x 4] | | | | |
| 1 | Age | 11 | 0 | 0 | 0 |
| 2 | Avr | 17 | 0 | 3 | 5,28 |
| 3 | Bil | 8 | 0 | 1 | 0 |
| 4 | Bod | 25 | 2 | 5 | 1,35 |
| 5 | Cha | 3 | 0 | 0 | 0 |
| 6 | Chu | 10 | 0 | 0 | 0 |
| 7 | Dob | 0 | 0 | 0 | 0,3 |
| 8 | Dor | 5 | 0 | 0 | 0 |
| 9 | Ere | 11 | 0 | 0 | 0,9 |
| 10 | Evl | 3 | 1 | 1 | 0 |
| 11 | Fil | 10 | 0 | 0 | 0,7 |
| 12 | Gol | 6 | 0 | 0 | 0,4 |
| 13 | Gre | 3 | 0 | 0 | 0,05 |
| 14 | Gry | 3 | 0 | 0 | 0,52 |
| 15 | Hak | 17 | 0 | 8 | 0 |
| 16 | Kar | 8 | 1 | 2 | 0,54 |
| 17 | KoA | 14 | 0 | 0 | 0,86 |
| 18 | Kob | 9 | 0 | 0 | 0,15 |
| 19 | Las | 19 | 0 | 0 | 0,06 |
| 20 | Lem | 19 | 1 | 1 | 0,67 |
| 21 | Mas | 5 | 0 | 1 | 1,15 |
| 22 | Ner | 13 | 0 | 0 | 0 |
| 23 | Nev | 6 | 0 | 2 | 0 |
| 24 | Pol | 7 | 0 | 0 | 0 |
| 25 | Rub | 7 | 2 | 3 | 0,15 |
| 26 | She | 4 | 0 | 0 | 0 |
| 27 | Sok | 8 | 0 | 0 | 0 |
| 28 | Sto | 3 | 0 | 0 | 0 |
| 29 | TkV | 7 | 0 | 0 | 0,15 |
| 30 | Vin | 0 | 0 | 0 | 0,23 |

**Figure 3.** Matrix $Rp$ [$30 \times 4$] contains the academic achievements' statistics (divided into 4 categories) from 30 staff members of the university. This matrix will be used as an example for demonstrating analytics suggested in this paper

## *Key atomic procedure 1: Administrative Ranking (based on Administrative Value System)*

Assume that administration of the organization (subject of assessment) publishes its "value system" as an instrument of transparent assessment of the academic personnel. Administrative value system is a vector $AVSa$ (<u>A</u>cademic <u>V</u>alue <u>S</u>ystem, <u>a</u>dministration) or a row [$1 \times n$], which indicates the value of relative importance administration gives for each academic achievement of a certain category. The vector is normalized as follows: $\sum_{i=1}^{n} AVSa(i) = 1$. For example, $AVSa(3) = 0.15$ would mean that the administration considers the relative importance of the 3rd category of achievements as 15% of the overall importance.

$AR$ (Administration, Ranking) is a vector (a row) $[1 \times m]$, which indicates the assessment score of each of $m$ staff members following the administrative value system. This vector (aka $[1 \times m]$ matrix) is computed as multiplication of the administrative value system vector (i.e., $[1 \times n]$ matrix) with the transpose matrix $[n \times m]$ of the academic results of the staff members as follows:

$$AR = AVSa \times Rp^T. \qquad (1)$$

Formula (1) uses matrix multiplication operation "×", which is defined as follows: If $A$ is an $[r \times p]$ matrix and $B$ is an $[p \times q]$ matrix, then the matrix product, $C = A \times B$, is defined to be the $[r \times p]$ matrix computed as follows: $\forall i (i = \overline{1,r}), \forall j (j = \overline{1,q}), C(i,j) = \sum_{k=1}^{p} A(i,k) \cdot B(k,j)$. Notice that the symbol "×" is also used throughout the paper to specify the size of matrixes [rows × columns].

Formula (1) also uses the transpose of a matrix operation, which is defined as follows: If $A$ is an $[r \times p]$ matrix, then the transpose matrix, $D = A^T$, is defined to be the $[p \times r]$ matrix computed as follows: $\forall i (i = \overline{1,p}), \forall j (j = \overline{1,r}), D(i,j) = A^T(i,j) = A(j,i)$.

### Key atomic procedure 2: Personnel Self-Ranking (based on Personnel Value System)

Assume that sometimes various types of a self-assessment will be applied within the organization and, therefore, personnel in such cases will be both subject and object of assessment. Each staff member of the organization publishes its personal "value system" as a transparent instrument for the self-assessment, and all individual value systems are stored as matrices. The value system of the personnel is a $[m \times n]$ matrix named $AVSp$ (Academic Value System, personnel), which indicates the value of relative importance that staff member (matrix row) gives for each academic achievement of a certain category (matrix column). The vector is normalized as follows: $\forall j(j = \overline{1,m}), \sum_{i=1}^{n} AVSp(j,i) = 1$. For example, $AVSp(5,2) = 0.3$ would mean that person number 5 considers the relative importance of the 2nd category of achievements as 30% of the overall importance.

$PR$ (Personnel, Self-Ranking) is a $[m \times m]$ square matrix, which indicates the assessment score of each of $m$ staff members given by following the personal value system of each of $m$ staff members. This matrix is computed as multiplication of the personnel value system matrix $[m \times n]$ with the transpose $[n \times m]$ matrix of the academic results of the staff members as follows:

$$PR = AVSp \times Rp^T . \qquad (2)$$

For demonstrating the experiments with the analytics suggested in this paper, we took the cuts from the real value systems (both from the administration and from the personnel) used in the academic organization (Figure 4). Here, we have four categories of achievements (as introduced in Figure 1) and each value system indicates relative importance of each category (i.e., relative weight of each quality criterion). Figure 4(a) contains vector $AVSa$ as an administrative value system and Figure 4(b) contains matrix $AVSp$ as set of personal value systems from each of 30 chosen staff members.

### *Key atomic procedure 3: Normalization of Rankings*

Values from AR and PR can be normalized to indicate not only the score but also the share of each staff member contribution in the total collective contribution regarding the value system.

We define normalization operation as follows. If $A$ is an $[r \times p]$ matrix, then the normalized matrix, $\|A\|$, is defined to be the $[r \times p]$ matrix computed as follows:

$$\forall i(i = \overline{1,r}), \forall j(j = \overline{1,p}), \|A\|(i,j) = \frac{A(i,j)}{\sum_{k=1}^{p} A(i,k)} . \qquad (3)$$

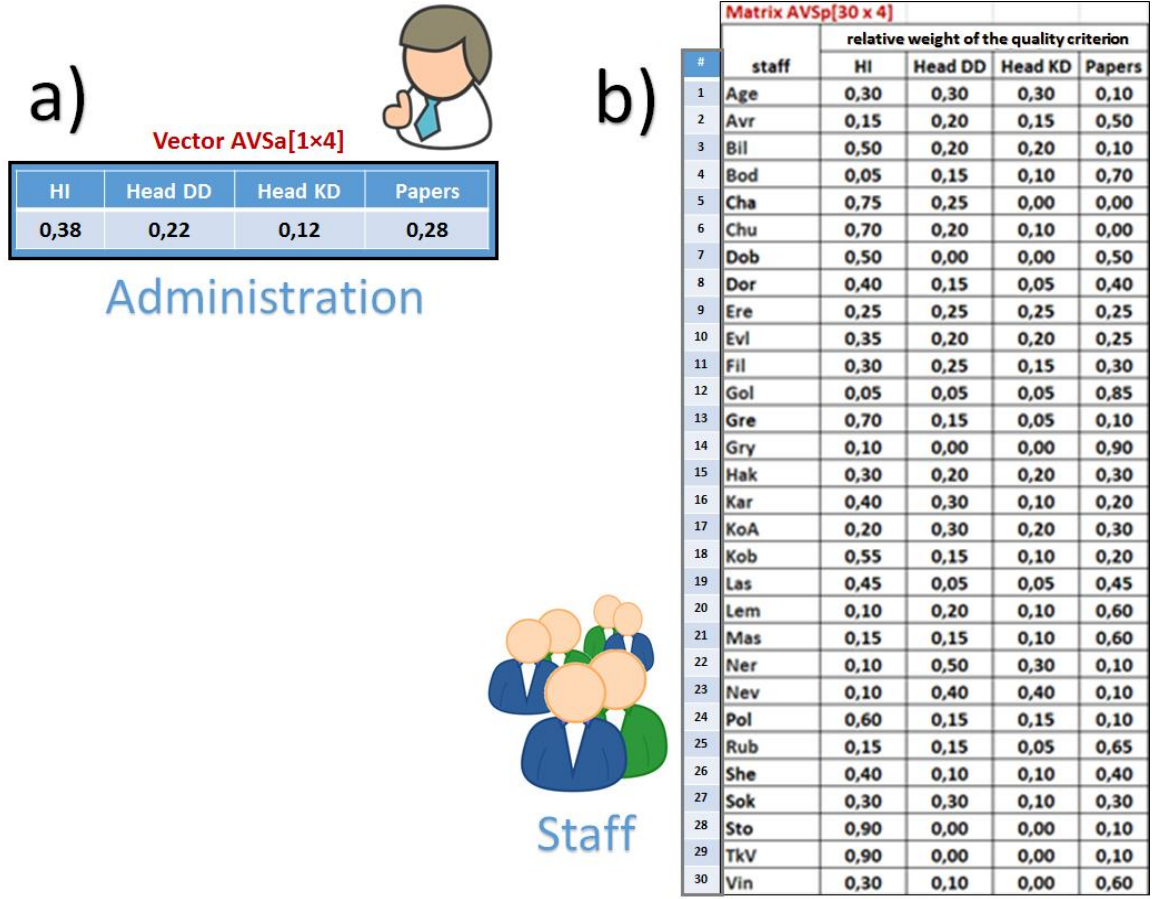

**Figure 4.** Value systems used in the experiments: (a) administrative value system or vector $AVSa$ $[1 \times 4]$; (b) value systems of the personnel or matrix $AVSp$ $[30 \times 4]$

Therefore, $AR$ as a vector-column or a $[1 \times m]$ matrix after the normalization will become the $\|AR\|$ matrix, which can be interpreted as the $AA$ (Administrative Assessment) ranking list computed as follows:

$$\forall j(j = \overline{1,m}), AA(j) = \|AR\|(1,j) = \frac{AR(1,j)}{\sum_{k=1}^{m} AR(1,k)} \ . \qquad (4)$$

$PR$ as a square $[m \times m]$ matrix will become (after the normalization) the $\|PR\|$ matrix computed as follows:

$$\forall i(i = \overline{1,m}), \forall j(j = \overline{1,m}), \|PR\|(i,j) = \frac{PR(i,j)}{\sum_{k=1}^{m} PR(i,k)} \ . \qquad (5)$$

For example, the value: $\|PR\|(2,5) = 0.012$ would mean that, following the value-system-driven "opinion" of staff member #2, staff member #5 contributed 1.2% to the common pool of the achieved collected score. Diagonal values, $\|PR\|(i,i)$, can be interpreted as self-evaluation of personal share to common contribution. If one needs representations of values for (4) or (5) in percent, then all the values of corresponding vector or matrix are multiplied by 100.

## 4. Compromise Group Decision-Making (Approaches and Analytics)

The importance of the participative decision-making for universities is increasing as their academic and financial autonomy grows. The transfer of some administrative powers to the employees transfers the management of the organization to the category of group-decision-making problems. The processes and the procedures, alongside the degree and the form of employee participation in decision-making may vary. To define possible approaches to the administration-staff-tradeoff-driven management, one need to define the rules for the compromises.

### 4.1. Analytics of Representative Participation

The administration alone cannot cover all the areas of expertise in the academic environment. Therefore, in problem areas related to developing organization (e.g., university), partial responsibility for decision-making can be delegated to the experts chosen from the personnel (representative participation). Concurrently, the administration retains the authority to form a strategic visions and principles for the final compilation of the results obtained.

#### 4.1.1. Case 1: Academic leagues
**Brief:**
- Evaluate and cluster the personnel into several groups (leagues) using administrative value system as an instrument and nominate the best performers within each league to lead the league;
- Re-evaluate each league separately using the value system of the corresponding leader.

**Intuition behind the approach:**

Academic staff are heterogeneous in their value creation capability; thus, optimal use of academic human resources within an organization requires a differentiated approach regarding this heterogeneity.

**Procedures**

1) Before the start of the assessment (ranking) procedure, the following are recorded and made public:

    a) "The rules of the game," which explain the ranking and decision-making algorithm based on the ranking results;

    b) Administrative value system (vector $AVSa$);

    c) Individual value systems of personnel (matrix $AVSp$).

2) During the assessment procedure, the achievements of each employee are recorded, verified, and made public, which allows, upon completion of the procedure, to form a matrix of personnel academic results (matrix $Rp$).

3) The primary rating of personnel achievements is performed following the administrative value system (see key atomic procedure 1); hence, the administrative ranking vector $AR$ is computed.

4) The resulting ranking list (i.e., vector $AR$ represented in descending order) is divided into three groups of equal size: the Major League (Seniors), the First League (Middles), and the Second League (Juniors).

5) In each group, its leader is nominated and this would be an employee with a top position in this group (top-Senior, top-Middle, or top-Junior).

6) Since now, the value system for each group re-evaluation will be the value system of its leader, i.e., the value system of the leader becomes an administrative value system within the league. If the leader is the $i$-th person within the organization than his/her value system (i.e., row $i$ in the matrix $AVSp$) will act as a vector $AVSa$ locally applicable within this particular group (league).

7) Each group is re-assessed and reordered following the value system of the group leader. This means that Steps 1–3 of this procedure are repeated but now within each league (as a smaller organization with its administration) separately.

8) The final staff ranking comprises three segments: the Major League ranking list (computed following the top-Senior value system), then the First League ranking list (computed following the top-Middle value system), and the Second League ranking list (computed following the top-Junior value system).

## Description of the experiment (usage example)

We experimented with the $[m \times n]$ (particularly $[30 \times 4]$) matrix $Rp$ as an input (Figure 3). In the experiment, we also used the administrative value system, i.e., vector $AVSa$ $[1 \times 4]$ from Figure 4(a) and value systems of the personnel, i.e., matrix $AVSp$ $[30 \times 4]$ from Figure 4(b).

First, the assessment procedure is performed according to formula (1), which computes the score for each staff member based on actual achievements and the administrative value system as a metrics. Figure 5 shows the resulting scores.

| $AVSa \times Rp^T$ | | | | | | | | | | | | | | | | | | | | | | | | | | | | | | |
|---|---|---|---|---|---|---|---|---|---|---|---|---|---|---|---|---|---|---|---|---|---|---|---|---|---|---|---|---|---|---|
| \multicolumn{31}{l}{Results of assessment using administrative value system} |
| Age | Avr | Bil | Bod | Cha | Chu | Dob | Dor | Ere | Evl | Fil | Gol | Gre | Gry | Hak | Kar | KoA | Kob | Las | Lem | Mas | Ner | Nev | Pol | Rub | She | Sok | Sto | TkV | Vin | |
| 4,18 | 8,30 | 3,16 | 10,92 | 1,14 | 3,80 | 0,08 | 1,90 | 4,43 | 1,48 | 4,00 | 2,39 | 1,15 | 1,29 | 7,42 | 3,65 | 5,56 | 3,46 | 7,24 | 7,75 | 2,34 | 4,94 | 2,52 | 2,66 | 3,50 | 1,52 | 3,04 | 1,14 | 2,70 | 0,06 | |

**Figure 5.** Scores of each staff member given by the administration for the assessed academic achievements in the experiment

Regarding the ordered scores, 10 best ones are nominated to the group (academic league) "Seniors"; the next 10 best ones will be in the group "Middles," and the last 10 go to the group "Juniors," as shown in Figure 6. In each of the three leagues, the leaders are determined by the top position (highest score within the group), i.e., top-Senior employee is "Bod," top-Middle employee is "Chu," and top-Junior employee is "Mas."

The final stage of ranking is reassessment of results within each league separately following the value systems of its leaders. This procedure does not change the content of each league (the recruited employees within the leagues remain unchanged), but induces a change in the ranking scores and corresponding position within each league (Figure 6).

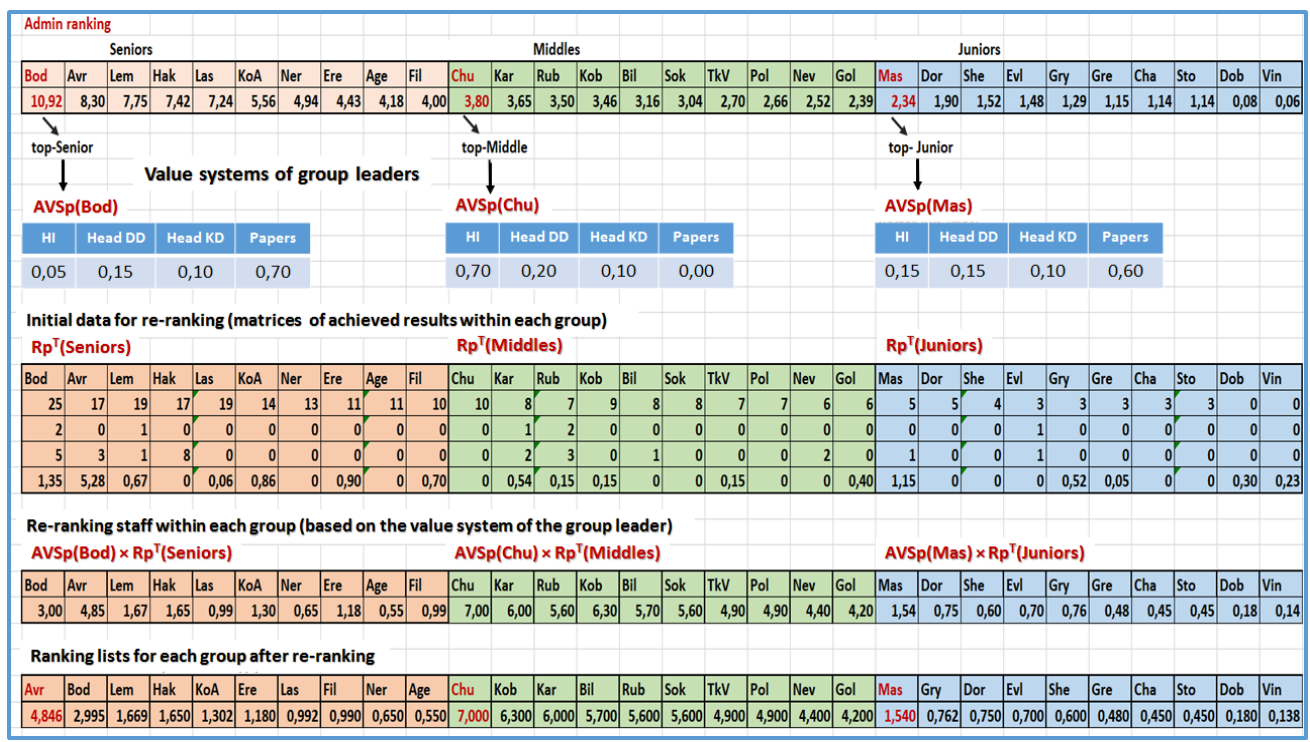

**Figure 6.** Reassessment procedure within the leagues in the experiment

## Observations and conclusive remarks

This kind of rating procedure has been practiced in NURE annually since 2017, allowing us to make several observations. Dividing the academic staff into the leagues concentrates competition on the local levels available to each employee. This keeps the motivation for gradual improvement regardless of the results already achieved by the employee. Competition at a simpler level serves to build the knowledge and experience needed to complete new sophisticated tasks. This forms the basis for the professional and career growth of young teachers. Simultaneously, the fulfillment of tasks of a higher level increases the status of an employee not only through improving his/her position in the ranking, but also through recognition of him/her as a part of more authoritative academic community (from the higher league up to the academic elite).

The promulgation of value systems plays a special role in this procedure. Individual decision-making criteria are not obvious in traditional mechanisms, as they are hidden behind the statistics. People can declare their priorities and publish their decisions; however, the true immediate reasons for each decision remain in the black box. This favors the use of double standards when people situationally adjust the declared general principles to the desirable decisions. For example, following the theory of rational behavior, a person seeks to optimize his/her efforts to obtain certain benefits. Therefore, it seems quite logical that the employee wants to also adjust the assessment criteria (weights used in ranking) so that the existing results of his/her work are assessed as high as possible. Concurrently, the employee may well be aware that such results are mediocre and do not induce the organization's development. Double standards may become a bad habit within the organization if the actual decision-making and the declared rules are not linked by transparent causation. Formalized and openly declared value systems not only make the decision-making process transparent, but also accurately explain a person's attitude toward different types of academic achievement. In this case, the desire to "adjust" the value system to suit your results (to "shift" the overall rating in your favor) will be offset by reputational risks. After all, the published system of values primarily characterizes the professional and team competencies of each individual employee.

It turns out that a meager value system may not be such a shameful thing if it is out of the public eye. However, it becomes undesirable if it is visible for everyone and available for public discussion. There is also one opposite effect, i.e., the value system, in contrast to the results (which can be obtained only by spending time and effort), allows the employee to immediately get into an advanced social environment just because he/she thinks as an elite.

To obtain the reputational and incentive impact from the published value systems, it is important to consider the timing factor also. The moment of promulgation of value systems should come in advance, i.e., before the beginning of the assessment period (or in extreme cases—before the start of the procedure).

Special attention must be paid to the specific role of the group leaders. These are representatives of the academic environment. The use of their value systems creates an opportunity for solidarity among academic staff in the assessment procedure, making it more democratic. Also, each group leader meets (better than others within the group do) the requirements of the administration. This ambivalence characterizes group leaders as compromise figures whose opinions weigh on both staff and administration.

Re-ranking following the value system of leaders can induce interesting consequences. One of these situations is a change in the top position in the group. In our experiment (Figure 6), in the "Seniors" league, the first and second positions were reversed after the re-rating. This means that the leader ("Bod"), "chosen" by the administration, considers the achievements of another employee ("Avr") more significant than his own. This castling shows how the principle of honesty inherent in the procedure works: the disclosure of one's criteria for assessing academic quality allows a person to remain principled and truthful, adequately maintaining loyalty to his/her viewpoint in any (even not beneficial for himself) situations.

The results within the leagues provide valuable information for further HR analytics:
1) The difference between the leagues regarding the achievements' score characterizes the ratio of labor productivity of certain categories of workers, which is important for identifying the weak and strong characteristics of the academic capital. In our experiment (Figure 7), the major ("Seniors") league showed a 2.73-fold gap in the results (from a maximum of 10.92 score to a minimum of 4.00 score), which, unfortunately, diagnoses a qualitative heterogeneity in the major league. Obviously, it should be admitted that the academic elite of the diagnosed university is limited to no more than the first half of the top-league rating list, while the second half of the major league regarding the quality of its added value is much closer to the Middles. In our opinion, this fact deserves much more attention from the administration than, for example, the gap in the results of the second league by 36.4 times. Given the extremely small number of achievements in the second ("Juniors") league (from 2.34 to 0.06 points), this gap lacks a fundamental effect on the general situation. It simply reflects a group of employees with an insignificant (zero or tending to zero) added value. The first ("Middles") league, in contrast, has a high density of indicators with a minimum performance gap of 1.59 times (from 3.8 to 2.39). This can be positively interpreted only with sufficiently large absolute values of the assessment scores. Otherwise, it is highly likely that we are exploring an established informal culture of low-quality work.

2) The directions of the vectors of individual achievements characterize the competitive diversity of the league. Employees can compete using just one criterion (for example, the Hirsch-index), which is fraught with the danger of imbalance in the development. Otherwise, they can compete using multiple criteria (someone generates more publications, while others focus on dissertation leadership, etc.), which makes the organization's development more sustainable.

3) Evaluation can be complicated by the appearance within the league (or, even worse, on the border of the leagues) groups with the same score of the result. In this case, a well-thought-out and agreed in advance formal procedure for differentiating such results is required. If this situation is aggravated by identical sets of achievements, this already testifies to the primitiveness of competition in this segment and requires a revision of personnel requirements.

It is convenient to interpret the results of the ranking of academic personnel as a kind of "inventory" of academic capital, based on the results of which it is possible to make not only operational decisions (such as bonuses or filling the vacant positions), but also strategic ones.

The major league ("Seniors") contains the most valuable academic resource that produces the organization's greatest added value. This resource is critical for the effectiveness of the university; therefore, it must be preserved in every possible way without interfering with its growth, encouraging the creation of scientific schools, and promoting it in the external academic environment as representatives of the university. It is also important that these are the leading professional experts of the university having a significant expert view of developing sciences and industries. Therefore, this group must have the greatest share in the decision-making within the organization.

The first league ("Middles") can be seen as the academic potential of the organization. Employees in this group are currently producing less outstanding results. However, they have growth prospects since they have already demonstrated their ability to develop, "stepping over" the second ("Juniors") league. Since the first league contains actively developing personnel, the administration needs to allocate resources to support it. The main barriers on this path are barriers to access: research resources (equipment, laboratories, libraries, etc.); scientific communications (participation in conferences, publications, scientific communities, etc.); research interaction (attraction to work in more powerful scientific groups, participation in industrial projects, etc.); and mastering new relevant knowledge and developing new skills (the possibility of high-quality LLL, certification, etc.). Even by providing selective support to the employees of the first league in these areas, it is possible to significantly increase the chances of qualitative growth of the added value of the university.

The second league ("Juniors") brings the organization the lowest added value, caused by two reasons: either low abilities or the undiscovered potential of each employee. It is also necessary to encourage accelerated professional growth. This task does not require close attention of the administration. It can be addressed by encouraging mentoring and internal communication at the departmental level. Juniors' potential is unclear, as they lack success story yet. This league creates the academic reserve of the organization for two processes: for development in the future (for example, regarding young teachers) and for "dumping ballast," regarding necessary staff reductions.

### 4.1.2. Case 2: Social lift (modification of academic leagues)

**Brief**

Social lift enables rotation of the status positions of academic staff between the academic leagues to accelerate social mobility within the university.

**Intuition behind the approach**

Development of the organization requires the acceleration of "metabolic" processes. Social lift provides a simplified procedure for social mobility (that is, the movement of an employee from one status position to another) based on demonstrated academic ability. The principle of forming football leagues is acceptable for creating a social lift.

**Procedures**

1) Based on academic leagues formed following the administrative system of values and reassessment (re-ranking) following the value systems of leaders, each employee's position in the overall ranking list is determined (see description of Case 1).
2) Three first positions and three last positions in each league are determined.
3) The last three positions in the higher league are swapped with the first three positions in the lower league, inducing the final ranking list *SLR* (Social Lift Ranking).

**Description of the experiment (usage example)**

To show the working principle of social lift at the university, we used, as an input, the illustrative sample and the results of the previous case (Figure 6). Therefore, we accept as initial information: the academic leagues derived from the administrative rankings and the reassessment scores of the employees obtained based on the value systems of the group leaders. Thus, we determine the serial number of each employee (Figure 7).

**Figure 7.** Illustrating the work of social lift in the experiment

## Observations and conclusive remarks

While the social lift is a reward for the best results in one's league (upward social mobility), it is a sanction for the worst results in one's league (downward social mobility). This technique is an efficient instrument for staff motivation because:

- It does not require additional financial or human resources from the administration (such as bonuses or new full-time positions);
- It is easily explainable due to the transparency of the procedure and common understanding of the popular football leagues' organizational principle by almost all segments of the population;
- Its results are temporary: if the employees who have received a credit of trust from the administration for upward social mobility do not justify it later on, then in the next round of rating, they will return to the lower league.

It is important to say that grouping staff into leagues based on their achievement scores (also known as similarly performed persons) is just one reasonable way to enable group-wise adaptive analytics within the organization. Conceptually different way of grouping would be clustering of the personnel based on their personal academic value systems (matrix $AVSp$), meaning that people with similar views toward assessment of the academic achievements (weights for different attributes) will join the same cluster, also known as like-minded persons. Further ranking analytics will be applied separately within each cluster the same way as described above for the leagues. Social mobility (horizontal) between the groups (instead of social lift used for leagues) would be exchanging from time to time typical (closer to mean) representatives of the groups and observing the culture interpenetration effect within the clusters. If to apply the 3-mean clustering algorithm to the 4-dimensional space of academic value systems of the personnel within our experiment, then (in addition to leagues), we will get the following three clusters of the like-minded employees (also the two most typical representatives of each cluster are discovered as potential exchange candidates to two other groups):

CLUSTER 1: Avr; Bod; Dob; Gol; Gry; Las; Lem; Mas; Rub; and Vin. (Exchange candidates: Mas and Rub).

CLUSTER 2: Age; Bil; Cha; Chu; Gre; Ner; Nev; Pol; Sto; and TkV. (Exchange candidates: Bil and Pol).

CLUSTER 3: Dor; Ere; Evl; Fil; Hak; Kar; KoA; Kob; She; and Sok. (Exchange candidates: Evl and Fil).

### 4.2. Analytics of the Participative Leadership

For a university, *participative leadership* means that the rector and the university board delegate certain powers to all the academic staff. Since this assumes that all the members of the academic community participate in making certain decisions, such compromises can be considered democratic.

#### 4.2.1. Case 3: Group self-assessment of the academic community

**Brief**

Group self-assessment is a democratic assessment of an employee by all the representatives of the academic community, when "everyone evaluates everyone" based on the individual value systems.

**Intuition behind the approach**

Each employee acts both as an object and as a subject of assessment. An assumption (a compromise) within the approach is an equal importance given to the opinion of each employee during the assessment. The administration does not participate in the assessment and ranking processes.

**Procedures**

1) Preparation stage:
   - Prior to the commencement of the assessment procedure, the "rules of the game" and individual value systems of personnel for assessing academic results (matrix $AVSp$) are collected and published.
   - During the assessment procedure, the achievements are collected, recorded, verified, and made public as matrix of personnel academic results (matrix $Rp$).

2) Computing the "democratic self-evaluation" ranking list as group self-assessment of the academic community.
   - Key atomic procedure 2 (Personnel Self-Ranking) is applied, i.e., individual ranking lists are computed following the value systems of each employee, see formula (2). The resulting $[m \times m]$ square matrix, $PR = AVSp \times Rp^T$, shows how each evaluates each. In the $PR$ matrix, horizontally, employees act as experts (the subject of assessment); vertically, the same employees act as the evaluated ones (the object of assessment). Therefore, individual ranking lists are the rows of the matrix, each element of which is the evaluation score of a particular employee (whose identifier is the column title) through the eyes of the particular expert (whose identifier is the row title).
   - Individual ranking lists are normalized following the key atomic procedure 3. The resulting matrix $\|PR\|$, computed according to the formula (5), shows the share of each employee in the results of the university according to each expert assessment.

- Finally, the democratic (group) assessment of each employee by the academic community is calculated as a simple average assessment of this employee by all the experts; therefore, we compute the $DA$ (Democratic Assessment) vector $[1 \times m]$ as follows:

$$\forall i (i = \overline{1,m}), DA(i) = \frac{\sum_{k=1}^{m} \|PR\|(k,i)}{m} \quad (6)$$

- The final democratic self-evaluation ranking list of the academic community will be the descending ordered $DA$ vector.

**Comments on the procedure**

Ranking lists without normalization (in particular, rows of the $PR$ matrix) should not be compared since the aggregate scores of the result (the sum of the values by rows) will differ for each expert. However, it is possible to compare these total scores themselves: they reflect the opinion of this expert about the total score of the university's performance. Thus, one can single out the most optimistic view of the university's achievements (experts with the maximum total amount) and the most pessimistic (with the minimum total amount). Also, for the comparability of the estimates obtained, the individual ranking lists are passed through the standardization (normalization) procedure, where the entire sum of the university's achievements is equal to 100%. Such normalized list shows the share of each individual employee to the common contribution. The diagonal of the matrix, $\|PR\|$, of individual normalized ranking lists reflects the self-esteem of each employee: that is, how each employee evaluates own results (share) following his/her own value system. Comparing two values (i.e., (a) how an employee evaluates oneself (diagonal value) with (b) how (in average) all others evaluate the same employee) gives a good indicator of the humility, modesty, objectivity, and self-criticism of the employee or, otherwise, indicates some overconfidence.

**Description of the experiment (usage example)**

As in previous case, we experimented with the $[m \times n]$ (particularly $[30 \times 4]$) matrix of the achievements $Rp$ (Figure 3), and we used the value systems matrix $AVSp$ $[30 \times 4]$ from Figure 4(b).

We applied personnel self-ranking procedure according to formula (2) and obtained individual ranking lists collected to the $[m \times m]$ square matrix $PR$, which is shown in Figure 8.

**Figure 8.** Self-assessment matrix $PR$ before normalization in the experiment

Matrix $PR$ is an intermediate stage within the democratic self-assessment procedure, as it does not allow comparison of matrix elements in columns. Due to the different aggregate scores of achievements provided by each individual assessment, we cannot determine which expert ranked a given employee higher or lower. However, this matrix can be interesting for comparing the line-by-line aggregate scores. These scores indicate the overall assessment of the university-as-whole achievements provided by each expert. In our experiment, the scatter of the aggregate scores is quite large: from the optimistic 236.25 points (according to "Sto" and "TkV" employees) to the pessimistic 26.19 (according to the "Gol" employee).

| % | Staff members as an object of assessment | | | | | | | | | | | | | | | | | | | | | | | | | | | | | | | Σ |
|---|---|---|---|---|---|---|---|---|---|---|---|---|---|---|---|---|---|---|---|---|---|---|---|---|---|---|---|---|---|---|---|---|
| | | Age | Avr | Bil | Bod | Cha | Chu | Dob | Dor | Ere | Evl | Fil | Gol | Gre | Gry | Hak | Kar | KoA | Kob | Las | Lem | Mas | Ner | Nev | Pol | Rub | She | Sok | Sto | TkV | Vin | |
| Staff members as a subject of assessment (experts) | Age | 3,67 | 7,27 | 3,01 | 10,84 | 1,00 | 3,34 | 0,03 | 1,67 | 3,77 | 1,67 | 3,42 | 2,05 | 1,01 | 1,06 | 8,35 | 3,73 | 4,77 | 3,02 | 6,35 | 7,09 | 2,13 | 4,34 | 2,67 | 2,34 | 4,02 | 1,34 | 2,67 | 1,00 | 2,35 | 0,03 | 100 |
| | Avr | 3,21 | 10,99 | 2,63 | 10,86 | 0,88 | 2,92 | 0,29 | 1,46 | 4,09 | 1,56 | 3,60 | 2,14 | 0,93 | 1,38 | 7,31 | 3,84 | 4,93 | 2,78 | 5,61 | 6,89 | 2,87 | 3,80 | 2,34 | 2,05 | 3,85 | 1,17 | 2,34 | 0,88 | 2,19 | 0,22 | 100 |
| | Bil | 3,97 | 6,94 | 3,03 | 10,12 | 1,08 | 3,61 | 0,02 | 1,80 | 4,03 | 1,37 | 3,66 | 2,19 | 1,09 | 1,12 | 7,28 | 3,36 | 5,11 | 3,26 | 6,86 | 7,19 | 2,03 | 4,69 | 2,45 | 2,52 | 3,26 | 1,44 | 2,89 | 1,08 | 2,54 | 0,02 | 100 |
| | Bod | 2,10 | 18,48 | 1,91 | 11,42 | 0,57 | 1,91 | 0,80 | 0,95 | 4,50 | 1,53 | 3,78 | 2,21 | 0,71 | 1,96 | 6,29 | 4,30 | 4,97 | 2,12 | 3,78 | 6,36 | 4,40 | 2,48 | 1,91 | 1,33 | 4,02 | 0,76 | 1,53 | 0,57 | 1,74 | 0,61 | 100 |
| | Cha | 4,18 | 6,46 | 3,04 | 9,75 | 1,14 | 3,80 | 0,00 | 1,90 | 4,18 | 1,27 | 3,80 | 2,28 | 1,14 | 1,14 | 6,46 | 3,16 | 5,32 | 3,42 | 7,22 | 7,34 | 1,90 | 4,94 | 2,28 | 2,66 | 2,91 | 1,52 | 3,04 | 1,14 | 2,66 | 0,00 | 100 |
| | Chu | 4,12 | 6,53 | 3,05 | 9,85 | 1,12 | 3,75 | 0,00 | 1,87 | 4,12 | 1,28 | 3,75 | 2,25 | 1,12 | 1,12 | 6,80 | 3,21 | 5,25 | 3,37 | 7,12 | 7,28 | 1,93 | 4,87 | 2,36 | 2,62 | 3,00 | 1,50 | 3,00 | 1,12 | 2,62 | 0,00 | 100 |
| | Dob | 4,01 | 8,12 | 2,91 | 9,60 | 1,09 | 3,64 | 0,11 | 1,82 | 4,34 | 1,09 | 3,90 | 2,33 | 1,11 | 1,28 | 6,19 | 3,11 | 5,41 | 3,33 | 6,94 | 7,17 | 2,24 | 4,74 | 2,19 | 2,55 | 2,61 | 1,46 | 2,91 | 1,09 | 2,61 | 0,08 | 100 |
| | Dor | 3,92 | 8,08 | 2,90 | 9,89 | 1,07 | 3,57 | 0,11 | 1,78 | 4,24 | 1,25 | 3,82 | 2,28 | 1,09 | 1,26 | 6,42 | 3,27 | 5,30 | 3,26 | 6,80 | 7,19 | 2,24 | 4,64 | 2,23 | 2,50 | 2,95 | 1,43 | 2,85 | 1,07 | 2,55 | 0,08 | 100 |
| | Ere | 3,57 | 8,20 | 2,92 | 10,81 | 0,97 | 3,24 | 0,10 | 1,62 | 3,86 | 1,62 | 3,47 | 2,07 | 0,99 | 1,14 | 8,10 | 3,74 | 4,82 | 2,97 | 6,18 | 7,03 | 2,32 | 4,21 | 2,59 | 2,27 | 3,94 | 1,30 | 2,59 | 0,97 | 2,32 | 0,07 | 100 |
| | Evl | 3,79 | 7,75 | 2,96 | 10,33 | 1,03 | 3,45 | 0,07 | 1,72 | 4,01 | 1,43 | 3,62 | 2,17 | 1,05 | 1,16 | 7,44 | 3,48 | 5,04 | 3,14 | 6,57 | 7,11 | 2,20 | 4,48 | 2,46 | 2,41 | 3,44 | 1,38 | 2,76 | 1,03 | 2,45 | 0,06 | 100 |
| | Fil | 3,74 | 8,09 | 2,89 | 10,39 | 1,02 | 3,40 | 0,10 | 1,70 | 4,05 | 1,47 | 3,64 | 2,18 | 1,04 | 1,20 | 7,15 | 3,53 | 5,06 | 3,11 | 6,49 | 7,15 | 2,26 | 4,42 | 2,38 | 2,38 | 3,51 | 1,36 | 2,72 | 1,02 | 2,43 | 0,08 | 100 |
| | Gol | 2,10 | 20,95 | 1,72 | 10,49 | 0,57 | 1,91 | 0,97 | 0,95 | 5,02 | 0,95 | 4,18 | 2,44 | 0,73 | 2,26 | 4,77 | 3,85 | 5,46 | 2,20 | 3,82 | 6,18 | 4,88 | 2,48 | 1,53 | 1,34 | 2,78 | 0,76 | 1,53 | 0,57 | 1,82 | 0,75 | 100 |
| | Gre | 4,13 | 6,75 | 3,03 | 9,75 | 1,13 | 3,75 | 0,02 | 1,88 | 4,18 | 1,23 | 3,79 | 2,27 | 1,13 | 1,15 | 6,60 | 3,17 | 5,30 | 3,39 | 7,14 | 7,28 | 1,97 | 4,88 | 2,31 | 2,63 | 2,88 | 1,50 | 3,00 | 1,13 | 2,64 | 0,01 | 100 |
| | Gry | 2,88 | 16,88 | 2,09 | 9,72 | 0,79 | 2,62 | 0,71 | 1,31 | 5,00 | 0,79 | 4,27 | 2,51 | 0,90 | 2,01 | 4,45 | 3,37 | 5,69 | 2,71 | 5,11 | 6,55 | 4,02 | 3,40 | 1,57 | 1,83 | 2,19 | 1,05 | 2,09 | 0,79 | 2,19 | 0,54 | 100 |
| | Hak | 3,70 | 8,17 | 2,92 | 10,44 | 1,01 | 3,37 | 0,10 | 1,68 | 4,01 | 1,46 | 3,60 | 2,15 | 1,03 | 1,18 | 7,52 | 3,55 | 5,00 | 3,08 | 6,41 | 7,07 | 2,29 | 4,38 | 2,47 | 2,36 | 3,53 | 1,35 | 2,69 | 1,01 | 2,41 | 0,08 | 100 |
| | Kar | 3,93 | 7,29 | 2,95 | 10,16 | 1,07 | 3,57 | 0,05 | 1,79 | 4,09 | 1,43 | 3,70 | 2,22 | 1,08 | 1,17 | 6,79 | 3,40 | 5,16 | 3,24 | 6,80 | 7,27 | 2,08 | 4,65 | 2,32 | 2,50 | 3,33 | 1,43 | 2,86 | 1,07 | 2,53 | 0,04 | 100 |
| | KoA | 3,45 | 8,76 | 2,82 | 10,99 | 0,94 | 3,14 | 0,14 | 1,57 | 3,88 | 1,73 | 3,47 | 2,07 | 0,96 | 1,19 | 7,84 | 3,86 | 4,80 | 2,89 | 5,99 | 7,06 | 2,42 | 4,08 | 2,51 | 2,20 | 4,15 | 1,26 | 2,51 | 0,94 | 2,27 | 0,11 | 100 |
| | Kob | 4,03 | 7,14 | 3,00 | 9,88 | 1,10 | 3,67 | 0,04 | 1,83 | 4,15 | 1,27 | 3,76 | 2,25 | 1,11 | 1,17 | 6,77 | 3,24 | 5,25 | 3,32 | 6,98 | 7,22 | 2,05 | 4,77 | 2,33 | 2,57 | 2,99 | 1,47 | 2,93 | 1,10 | 2,59 | 0,03 | 100 |
| | Las | 3,95 | 8,13 | 2,92 | 9,75 | 1,08 | 3,59 | 0,11 | 1,80 | 4,28 | 1,16 | 3,85 | 2,30 | 1,10 | 1,27 | 6,43 | 3,19 | 5,34 | 3,29 | 6,85 | 7,15 | 2,25 | 4,67 | 2,24 | 2,52 | 2,77 | 1,44 | 2,88 | 1,08 | 2,57 | 0,08 | 100 |
| | Lem | 2,87 | 13,50 | 2,35 | 11,00 | 0,78 | 2,61 | 0,47 | 1,31 | 4,28 | 1,57 | 3,71 | 2,19 | 0,86 | 1,60 | 6,53 | 3,98 | 5,01 | 2,59 | 5,06 | 6,80 | 3,37 | 3,40 | 2,09 | 1,83 | 3,89 | 1,05 | 2,09 | 0,78 | 2,06 | 0,36 | 100 |
| | Mas | 3,24 | 11,81 | 2,55 | 10,51 | 0,88 | 2,94 | 0,35 | 1,47 | 4,30 | 1,37 | 3,77 | 2,24 | 0,94 | 1,49 | 6,57 | 3,68 | 5,13 | 2,82 | 5,66 | 6,87 | 3,02 | 3,83 | 2,16 | 2,06 | 3,41 | 1,18 | 2,35 | 0,88 | 2,24 | 0,27 | 100 |
| | Ner | 2,82 | 8,01 | 2,82 | 13,15 | 0,77 | 2,56 | 0,08 | 1,28 | 3,05 | 2,82 | 2,74 | 1,64 | 0,78 | 0,90 | 10,50 | 5,00 | 3,81 | 2,34 | 4,88 | 7,09 | 2,34 | 3,33 | 3,07 | 1,79 | 6,70 | 1,02 | 2,05 | 0,77 | 1,83 | 0,06 | 100 |
| | Nev | 2,68 | 8,35 | 2,92 | 13,24 | 0,73 | 2,44 | 0,07 | 1,22 | 2,90 | 2,68 | 2,61 | 1,56 | 0,74 | 0,86 | 11,94 | 5,00 | 3,62 | 2,23 | 4,64 | 6,74 | 2,47 | 3,17 | 3,41 | 1,71 | 6,61 | 0,97 | 1,95 | 0,73 | 1,74 | 0,06 | 100 |
| | Pol | 4,05 | 6,86 | 3,04 | 9,93 | 1,10 | 3,68 | 0,02 | 1,84 | 4,10 | 1,29 | 3,72 | 2,23 | 1,11 | 1,14 | 6,99 | 3,25 | 5,20 | 3,32 | 7,00 | 7,22 | 2,00 | 4,78 | 2,39 | 2,58 | 3,05 | 1,47 | 2,94 | 1,10 | 2,59 | 0,01 | 100 |
| | Rub | 3,28 | 12,19 | 2,49 | 10,29 | 0,89 | 2,98 | 0,39 | 1,49 | 4,44 | 1,29 | 3,89 | 2,31 | 0,96 | 1,57 | 5,86 | 3,58 | 5,29 | 2,88 | 5,74 | 6,93 | 3,08 | 3,88 | 1,99 | 2,09 | 3,18 | 1,19 | 2,39 | 0,89 | 2,28 | 0,30 | 100 |
| | She | 3,89 | 8,14 | 2,92 | 9,93 | 1,06 | 3,53 | 0,11 | 1,77 | 4,21 | 1,24 | 3,78 | 2,26 | 1,08 | 1,24 | 6,71 | 3,28 | 5,25 | 3,23 | 6,74 | 7,13 | 2,26 | 4,59 | 2,30 | 2,47 | 2,97 | 1,41 | 2,83 | 1,06 | 2,53 | 0,08 | 100 |
| | Sok | 3,79 | 8,01 | 2,87 | 10,33 | 1,03 | 3,44 | 0,10 | 1,72 | 4,10 | 1,49 | 3,68 | 2,20 | 1,05 | 1,21 | 6,77 | 3,51 | 5,12 | 3,15 | 6,56 | 7,23 | 2,23 | 4,48 | 2,30 | 2,41 | 3,49 | 1,38 | 2,75 | 1,03 | 2,46 | 0,08 | 100 |
| | Sto | 4,19 | 6,70 | 3,05 | 9,58 | 1,14 | 3,81 | 0,01 | 1,90 | 4,23 | 1,14 | 3,84 | 2,30 | 1,14 | 1,16 | 6,48 | 3,07 | 5,37 | 3,43 | 7,24 | 7,27 | 1,95 | 4,95 | 2,29 | 2,67 | 2,67 | 1,52 | 3,05 | 1,14 | 2,67 | 0,01 | 100 |
| | TkV | 4,19 | 6,70 | 3,05 | 9,58 | 1,14 | 3,81 | 0,01 | 1,90 | 4,23 | 1,14 | 3,84 | 2,30 | 1,14 | 1,16 | 6,48 | 3,07 | 5,37 | 3,43 | 7,24 | 7,27 | 1,95 | 4,95 | 2,29 | 2,67 | 2,67 | 1,52 | 3,05 | 1,14 | 2,67 | 0,01 | 100 |
| | Vin | 3,79 | 9,50 | 2,76 | 9,77 | 1,03 | 3,45 | 0,21 | 1,72 | 4,41 | 1,15 | 3,93 | 2,34 | 1,07 | 1,39 | 5,86 | 3,24 | 5,42 | 3,20 | 6,59 | 7,12 | 2,52 | 4,48 | 2,07 | 2,41 | 2,74 | 1,38 | 2,76 | 1,03 | 2,52 | 0,16 | 100 |
| AVERAGE | | 3,575 | 9,358 | 2,783 | 10,41 | 0,975 | 3,25 | 0,187 | 1,625 | 4,135 | 1,424 | 3,685 | 2,199 | 1,006 | 1,298 | 6,988 | 3,568 | 5,085 | 3,018 | 6,212 | 7,041 | 2,523 | 4,225 | 2,316 | 2,275 | 3,45 | 1,3 | 2,6 | 0,975 | 2,368 | 0,143 | 100 |

**Figure 9.** Self-assessment matrix //PR// after normalization in the experiment

Figure 9 shows the normalized matrix ‖PR‖, which is critical for comprehensive analytics. By analyzing the data by columns, you can see the assessment of a given employee by different experts. For example, let us check the leader of the administrative assessment "Bod" (see Figure 5), who has been ranked sufficiently high by all the employees as well (the values of the "Bod" column in the matrix):

- The minimal assessment of "Bod" contribution to the overall result was given by employees "Sto" and "TkV," but still they gave him 9.58% share of all the university (actually of the 30 employees' sample) achievements;
- "Bod" got the best score from "Nev," who estimated the contribution of "Bod" at 13.24%;
- "Bod" assessed himself at 11.42% (diagonal value), which slightly exceeds the average score given to him by others (10.41%);

- "Bod" himself, as an expert (line "Bod" in the matrix), gives the maximum score to his rival employee "Avr," evaluating his contribution at 18.48%. Let us remind you that this fact became decisive for the change of the leader during the re-rating in the major league (see Figure 6).

One can see that the academic community is unanimous in assessing achievements of "Bod"; however, there are employees who got quite diverse assessment scores from the community. For example, the contribution of the alternative leader "Avr" to the overall result of the team is assessed extremely ambiguously: from 6.46% according to "Cha" to 20.95% according to "Gol." Therefore, by analyzing the range of the personal assessment values, one can conclude about the cohesion (or disunity) in the views within a given academic environment.

Figure 10 shows the final democratic assessment of each employee (in our experiment) by the academic community (i.e., vector $DA$ and the corresponding ordered ranking list).

**Democratic (group) self-assessment results (vector $DA$)**

| Age | Avr | Bil | Bod | Cha | Chu | Dob | Dor | Ere | Evl | Fil | Gol | Gre | Gry | Hak | Kar | KoA | Kob | Las | Lem | Mas | Ner | Nev | Pol | Rub | She | Sok | Sto | TkV | Vin |
|---|---|---|---|---|---|---|---|---|---|---|---|---|---|---|---|---|---|---|---|---|---|---|---|---|---|---|---|---|---|
| 3,575 | 9,358 | 2,783 | 10,41 | 0,975 | 3,25 | 0,187 | 1,625 | 4,135 | 1,424 | 3,685 | 2,199 | 1,006 | 1,298 | 6,988 | 3,568 | 5,085 | 3,018 | 6,212 | 7,041 | 2,523 | 4,225 | 2,316 | 2,275 | 3,45 | 1,3 | 2,6 | 0,975 | 2,368 | 0,143 |

**Democratic (group) self-assessment ranking list (ordered vector $DA$)**

| Bod | Avr | Lem | Hak | Las | KoA | Ner | Ere | Fil | Age | Kar | Rub | Chu | Kob | Bil | Sok | Mas | TkV | Nev | Pol | Gol | Dor | Evl | She | Gry | Gre | Cha | Sto | Dob | Vin |
|---|---|---|---|---|---|---|---|---|---|---|---|---|---|---|---|---|---|---|---|---|---|---|---|---|---|---|---|---|---|
| 10,4 | 9,4 | 7,0 | 7,0 | 6,2 | 5,1 | 4,2 | 4,1 | 3,7 | 3,6 | 3,6 | 3,5 | 3,2 | 3,0 | 2,8 | 2,6 | 2,5 | 2,4 | 2,3 | 2,3 | 2,2 | 1,6 | 1,4 | 1,3 | 1,3 | 1,0 | 1,0 | 1,0 | 0,2 | 0,1 |
| 1 | 2 | 3 | 4 | 5 | 6 | 7 | 8 | 9 | 10 | 11 | 12 | 13 | 14 | 15 | 16 | 17 | 18 | 19 | 20 | 21 | 22 | 23 | 24 | 25 | 26 | 27 | 28 | 29 | 30 |

**Figure 10.** Results of the democratic self-assessment in the experiment

## Observations and conclusive remarks

Comparative analysis of ranking lists obtained by different procedures is also useful for an organization (see Figure 11). Observably, the assessment of outsiders (the last five positions) coincided in all three ranking options. This is completely logical as we are considering a lack of effective contribution from these employees. In such situation, the publication of different ranking lists plays an important motivational role. After all, if an outsider reveals his/her low position only in the administrative ranking list, he/she may consider it as an one-sidedness or a bias of the administrative vision. However, a similar position in the democratic self-assessment ranking list confirms that the entire team also considers the employee's contribution to be insufficient. Complaints about unproductive work, reinforced by double arguments, can become a more powerful incentive for an employee to reconsider their academic activities.

**Figure 11.** Comparing the results of three different ranking procedures in the experiment

While comparing the overall ranking results in the experiment, one can notice that the administration and the academic community are also unanimous in their assessments of the most productive workers. Their opinions differ in the middle of the rankings. However, the opinions of the leaders (those who achieved the best results in their subgroups or leagues) and the "social lift" rankings differ much from the results of both administrative and democratic assessments. If the few voices of the leaders "dissolve" in the average votes of the rest (less advanced) team, then such ranking results will not play a stimulating role in developing the university. Therefore, the philosophy of the "democratic self-evaluation" can be useful when:

- The strategic vision of the employees, on average, exceeds the strategic vision of the administration;
- The entire academic community thinks quite progressively, ensuring the leadership of the university in the external environment.

In other cases, the individual ranking lists are rather of analytical interest for understanding the internal processes in the academic environment of the university.

## 4.2.2. Case 4: Group self-assessment with the administration-staff compromise (weighted democracy)

**Brief**

Like the previous case, this is a group self-assessment or democratic self-evaluation of staff (based on individual value systems). However, here, the administration introduces a single factor of influence as a kind of staff-administration compromise: the administration weighs each employee following its value system and requires weighted aggregation of opinions during democratic self-assessment procedure.

**Intuition behind the approach**

Each employee acts both as an object and as a subject of assessment. The opinions of employees have different significance for the organization. The opinions of those who create more added value deserve more attention when assessing achievements at the organization. The administration determines the weight of the opinion of each employee in the self-assessment process.

**Procedures**

1) Preparation stage:
    - Prior to the start of the assessment procedure, the "rules of the game," the administrative value system (Vector $AVSa$) and the individual value systems of the personnel (Matrix $AVSp$) are collected and published.
    - During the assessment procedure, the achievements are recorded, verified, and made public, as a matrix of personnel academic results (Matrix $Rp$) is formed.
2) Administrative ranking:
    - An administrative ranking list (Vector $AR\ [1 \times m]$) is created according to formula (1).

- The administrative ranking list is normalized to the Vector $AA$ (which is $\|AR\|$) according to formula (4), representing the share of each employee in the overall result through the eyes of the administration.

3) Individual ranking:

- Individual rankings (Matrix $PR\ [m \times m]$) are computed based on the personal value systems of each employee according to formula (2), where each one evaluates each one.
- Individual rankings are normalized (Matrix $\|PR\|$), reflecting the share of each employee in the total result according to each expert.

4) Compromise ranking:

The compromise scores (Vector $CR$ (<u>C</u>ompromise Administration-Staff, <u>R</u>anking) as $[1 \times m]$ one-row matrix) for each employee are calculated as the average score of the employee by all the experts (individual standardized rating lists) by considering the weight of the opinion of each expert according to the administration:

$$CR \ = \ \|AR\| \ \times \|PR\| \ . \qquad (7)$$

**Comments on the procedure**

Vector $AR$ allows us to see through the eyes of the administration what part of the achievements each employee brought to the total results of the university. This is the key information for obtaining the compromise solution between the administration and the academic staff. The principle is simple: the more result you have brought to the university, the more would be the weight of your expertise in future decision process. By this way, an employee's performance is linked to his/her voting rights. Therefore, while the vector $AR$ serves as one of the options for ranking the personnel solely by the administration, it gives the weight of each employee in the compromise decision process as a kind of expert status assigned to the employee by the administration.

## Description of the experiment (usage example)

As in previous case, we experimented with the $[m \times n]$ (i.e., $[30 \times 4]$) matrix of the achievements $Rp$ (Figure 3), and we also used the value system of the administration, i.e., vector $AVSa$ $[1 \times 4]$ from Figure 4(a), and the value systems of the personnel, i.e., matrix $AVSp$ $[30 \times 4]$ from Figure 4(b).

Figure 12 depicts the results of the administration ranking, normalization (Vector $AA = \|AR\|$), and the final ranking list (following the administration). The normalized matrix, $\|PR\|$, has been already computed above (Figure 9).

| AVSa × Rpᵀ | Results of assessment using administrative value system (individual scores) |
|---|---|

| Age | Avr | Bil | Bod | Cha | Chu | Dob | Dor | Ere | Evl | Fil | Gol | Gre | Gry | Hak | Kar | KoA | Kob | Las | Lem | Mas | Ner | Nev | Pol | Rub | She | Sok | Sto | TkV | Vin |
|---|---|---|---|---|---|---|---|---|---|---|---|---|---|---|---|---|---|---|---|---|---|---|---|---|---|---|---|---|---|
| 4,18 | 8,30 | 3,16 | 10,92 | 1,14 | 3,80 | 0,08 | 1,90 | 4,43 | 1,48 | 4,00 | 2,39 | 1,15 | 1,29 | 7,42 | 3,65 | 5,56 | 3,46 | 7,24 | 7,75 | 2,34 | 4,94 | 2,52 | 2,66 | 3,50 | 1,52 | 3,04 | 1,14 | 2,70 | 0,06 |

| ‖AR‖=‖AVSa × Rpᵀ‖ | Normalized administrative assessments (individual weights %) |
|---|---|

| Age | Avr | Bil | Bod | Cha | Chu | Dob | Dor | Ere | Evl | Fil | Gol | Gre | Gry | Hak | Kar | KoA | Kob | Las | Lem | Mas | Ner | Nev | Pol | Rub | She | Sok | Sto | TkV | Vin |
|---|---|---|---|---|---|---|---|---|---|---|---|---|---|---|---|---|---|---|---|---|---|---|---|---|---|---|---|---|---|
| 3,88 | 7,70 | 2,93 | 10,13 | 1,06 | 3,53 | 0,08 | 1,76 | 4,11 | 1,37 | 3,71 | 2,22 | 1,07 | 1,19 | 6,89 | 3,39 | 5,16 | 3,21 | 6,72 | 7,19 | 2,17 | 4,59 | 2,34 | 2,47 | 3,25 | 1,41 | 2,82 | 1,06 | 2,51 | 0,06 |

**Ordered ranking list (administrative)**

| Bod | Avr | Lem | Hak | Las | KoA | Ner | Ere | Age | Fil | Chu | Kar | Rub | Kob | Bil | Sok | TkV | Pol | Nev | Gol | Mas | Dor | She | Evl | Gry | Gre | Cha | Sto | Dob | Vin |
|---|---|---|---|---|---|---|---|---|---|---|---|---|---|---|---|---|---|---|---|---|---|---|---|---|---|---|---|---|---|
| 10,13 | 7,70 | 7,19 | 6,89 | 6,72 | 5,16 | 4,59 | 4,11 | 3,88 | 3,71 | 3,53 | 3,39 | 3,25 | 3,21 | 2,93 | 2,82 | 2,51 | 2,47 | 2,34 | 2,22 | 2,17 | 1,76 | 1,41 | 1,37 | 1,19 | 1,07 | 1,06 | 1,06 | 0,08 | 0,06 |
| 1 | 2 | 3 | 4 | 5 | 6 | 7 | 8 | 9 | 10 | 11 | 12 | 13 | 14 | 15 | 16 | 17 | 18 | 19 | 20 | 21 | 22 | 23 | 24 | 25 | 26 | 27 | 28 | 29 | 30 |

**Figure 12.** Administrative ranking of the personnel (computing weight for everyone for potential collaborative decision-making)

To form a compromise estimate, we computed vector $CR$ $[1 \times m]$, each element of which is the average assessment of a given employee by all the team members, weighted by the importance of the opinion of each of them following the administration (Figure 13).

| Ranking list as compromise (personnel-administration) | CR = ‖AR‖ × ‖PR‖ = ‖AVSa × Rpᵀ‖ × ‖AVSp × Rpᵀ‖ |
|---|---|

| Age | Avr | Bil | Bod | Cha | Chu | Dob | Dor | Ere | Evl | Fil | Gol | Gre | Gry | Hak | Kar | KoA | Kob | Las | Lem | Mas | Ner | Nev | Pol | Rub | She | Sok | Sto | TkV | Vin |
|---|---|---|---|---|---|---|---|---|---|---|---|---|---|---|---|---|---|---|---|---|---|---|---|---|---|---|---|---|---|
| 3,39 | 10,11 | 2,71 | 10,66 | 0,93 | 3,09 | 0,24 | 1,54 | 4,10 | 1,51 | 3,64 | 2,17 | 0,97 | 1,33 | 7,15 | 3,72 | 5,00 | 2,90 | 5,91 | 6,98 | 2,69 | 4,01 | 2,33 | 2,16 | 3,69 | 1,23 | 2,47 | 0,93 | 2,28 | 0,18 |

**Ordered ranking list (compromise)**

| Bod | Avr | Hak | Lem | Las | KoA | Ere | Ner | Kar | Rub | Fil | Age | Chu | Kob | Bil | Mas | Sok | Nev | TkV | Gol | Pol | Dor | Evl | Gry | She | Gre | Cha | Sto | Dob | Vin |
|---|---|---|---|---|---|---|---|---|---|---|---|---|---|---|---|---|---|---|---|---|---|---|---|---|---|---|---|---|---|
| 10,66 | 10,11 | 7,15 | 6,98 | 5,91 | 5,00 | 4,10 | 4,01 | 3,72 | 3,69 | 3,64 | 3,39 | 3,09 | 2,90 | 2,71 | 2,69 | 2,47 | 2,33 | 2,28 | 2,17 | 2,16 | 1,54 | 1,51 | 1,33 | 1,23 | 0,97 | 0,93 | 0,93 | 0,24 | 0,18 |
| 1 | 2 | 3 | 4 | 5 | 6 | 7 | 8 | 9 | 10 | 11 | 12 | 13 | 14 | 15 | 16 | 17 | 18 | 19 | 20 | 21 | 22 | 23 | 24 | 25 | 26 | 27 | 28 | 29 | 30 |

**Figure 13.** Ranking list as a self-assessment compromise weighted by the administration (administration-staff compromise)

The resulting rating is a compromise of competencies:

- The academic environment comprises expert educators who can produce and assess the academic achievements;
- The administration determines the weight of the experts, i.e., it assesses everybody from the managers' viewpoint.

It may seem that such a "game of democracy" does not make sense because the administration will still gain the upper hand by giving the main role to "its" experts. However, it is not completely so. It is important that the administration gives preference to "ally" rather than "tamed" experts. There is a huge difference between these two concepts. If the rector chooses such an employee for the role of an expert, whose voice he can manipulate, this is a "tamed" expert. He gives those judgments that are needed not so much by the rector as an official, but by the person in the position of the rector. In the absence of transparency, this really creates a conducive environment for manipulation. In our case, it is not an ordinary person who can be persuaded, bribed, or intimidated to vote, but just the numbers—that is, the previously published system of values of this person. Emphatically, digitized and fixed decision-making criteria act as experts. If they are made public, the hidden manipulation of such "experts" becomes impossible. The principle, according to which an employee (or rather, corresponding system of values) becomes an expert "an ally for the administration," is not a compliance (which is technically impossible), but just the value of achievements (those achievements that promote the university and correspond to the strategic vision of not a human rector, but a rector position). However, the academic environment comprising many experts has enough power to influence the final assessments.

If one compares the administrative ranking as such with the ranking based on compromise, one can see the difference (Figure 14).

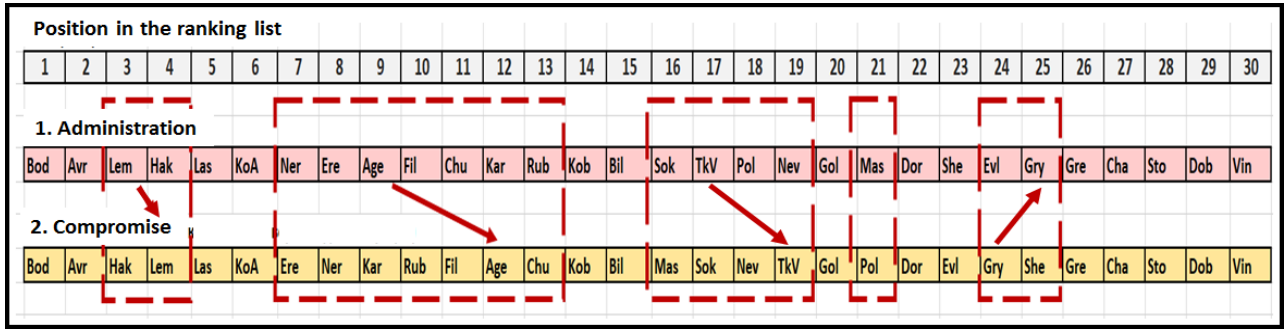

**Figure 14.** Comparing two ranking lists: (1) computed according to the purely administrative value system (Vector ||AR||); and (2) computed as the administration-staff compromise (Vector CR)

In our experiment, a rather large array of ranking positions does not match in the two compared versions (highlighted by red dotted line in Figure 14). Moreover, these discrepancies can also be small (as, for example, the rating of the "Lem" employee, whose position in the administrative ranking is 3rd, and in the compromise ranking, 4th) or significant (the position of the "Age" employee in the administrative ranking was 9th, and in the compromise, 12th).

The impression may arise that administrative participation in the compromise does not affect the final results because the academic environment is "pressing with its mass." However, this fear is wrong. By comparing the compromise ranking where the weight of experts is determined by the administration with the self-assessment of the academic environment, one can also see the discrepancy in some ranking positions (see Figure 15).

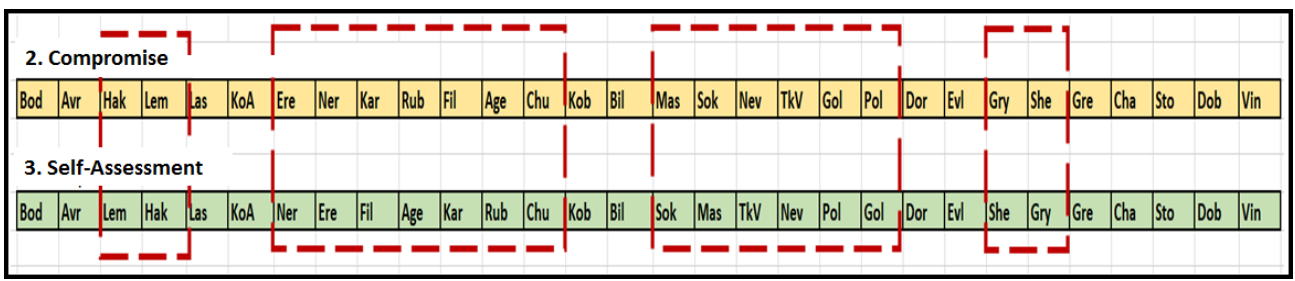

**Figure 15.** Comparing two ranking lists: (2) computed as the administration-staff compromise (Vector CR); and (3) computed as a purely democratic self-assessment following the personnel value system (Vector DA)

## Observations and conclusive remarks

Compromise ranking in which the weight of experts is determined by the administration is a good tool for the situations when:

- The academic environment of the university is sufficiently developed and strong, with achievements recognized in the academic community;
- The administration is also characterized as a strong management team with an undeniable track record of success;
- Development is characterized by a rather high uncertainty for the vectors of possible further development (a kind of bifurcation point for a university);
- There are serious differences in the opinions of the administration and the academic community.

Typically, such situations arise when industry leaders make pivotal decisions under the influence of changing environmental factors. A compromise will allow all the opinions to be considered: by giving up probably a little of something important by considering the majority of other important factors.

However, if in such a situation one of the players is weaker in his/her professional sphere, then the compromise solution will also be weakened and will stimulate less effective directions for the development. Also, this method is ineffective if a strict adherence to a certain development vector is necessary (for example, in a crisis management environment). Then, even very strong academic professionals may not consider organizational factors, and such a compromise will be vulnerable to further advancement in the external environment.

### 4.2.3. Case 5: Group self-assessment with the leader-staff compromise (expert democracy)

**Brief**

Like the previous case, this is a group self-assessment or democratic self-evaluation of staff (when everyone evaluates everyone), but the weight of the opinion of each employee is determined not by the administration, but by the leader ("winner").

**Intuition behind the approach**

- As in the previous case, each employee acts as an object and a subject of assessment; moreover, the opinions of employees have different significance for the organization;
- The opinions of those who create more added value deserve more attention when assessing academic achievements at the university;
- The administration does not interfere in the ranking process;
- The leader is conferred with the authority to determine the weight of the opinion of each employee (usually the winner of the group self-assessment).

**Procedure**

1) Preparation stage—same as for previous case.
2) Selecting the leader from the staff members (using one of already considered following options):
   - The first one in the ranking list created based on administration value system (key atomic procedure 1);
   - The first one after the reassessment within some subgroup or league (case 1);
   - The first one after the self-assessment procedure (case 3).
3) Individual ranking—same as in the previous case and resulted to the matrix $\|PR\|$ $[m \times m]$.
4) "Leader-Staff" Compromise ranking:
   - The row of matrix $\|PR\|$, which corresponds to the chosen leader, is represented as a vector $Lead$ $[1 \times m]$;
   - Computing compromise ranking list (Vector $CRL$ (<u>C</u>ompromise, <u>R</u>anking, <u>L</u>eaders-Staff) as $[1 \times m]$) as follows:

$$CRL = Lead \times \|PR\|. \qquad (8)$$

## Comments on the procedure

Notice that the administration does not directly participate in this ranking. All the responsibility for strategic vision of the priorities is transferred to the leader. That is why the choice of the leader plays a crucial role here and requires close attention. If a leader is identified among academic staff based on an administrative ranking, this gives reason to believe that the opinion of the administration has been considered in the ranking, albeit indirectly. If, however, the leader is the one who took the top position in the repeated rating within the leagues (case 1), then the influence of the administration will become even softer (like the first iteration in iterative algorithms).

## Description of the experiment (usage example)

As in previous case, we experimented with the $[m \times n]$ (i.e., $[30 \times 4]$) matrix of the achievements $Rp$ (Figure 3), and we also used the value systems of the personnel, i.e., matrix $AVSp$ $[30 \times 4]$ from Figure 4(b), and the results of the academic league experiment (case 1) from Figure 6.

For a comparative analysis, we will make three variants of such ranking with different leaders:

- In the first, the leader will present the top position of the administrative rating ("Bod" employee), and this is a premium class leader (according to the administration);
- In the second, the top position in the "Seniors" league ranking list from "Bod" (which is given to "Avr" employee), and this will be the leader of the premium class (according to the academic elite);
- In the third, the top position in the ranking list of the "Middles" league (i.e., employee "Chu" is the leader of that league).

The results of the compromise ranking with the choice of "Bod" as a leader are shown in Figure 16. The results of similar procedure with the choice of "Avr" as a leader are shown in Figure 17.

The results obtained (Figure 16 vs. Figure 17) demonstrate almost complete coincidence, except for the castling of the first two positions. The overlap in ranking positions could be attributed to similar value systems for the premium leaders: they both preferred the achievement of a category "Papers" more and the least, "HI."

**Leader = "Bod"**

| "Bod" assesses ("gives weights") to the "experts" (vector **Lead**, which is the row "Bod" in the Matrix ‖**PR**‖) | | | | | | | | | | | | | | | | | | | | | | | | | | | | | |
|---|---|---|---|---|---|---|---|---|---|---|---|---|---|---|---|---|---|---|---|---|---|---|---|---|---|---|---|---|---|
| Age | Avr | Bil | Bod | Cha | Chu | Dob | Dor | Ere | Evl | Fil | Gol | Gre | Gry | Hak | Kar | KoA | Kob | Las | Lem | Mas | Ner | Nev | Pol | Rub | She | Sok | Sto | TkV | Vin |
| 2,10 | 18,48 | 1,91 | 11,42 | 0,57 | 1,91 | 0,80 | 0,95 | 4,50 | 1,53 | 3,78 | 2,21 | 0,71 | 1,96 | 6,29 | 4,30 | 4,97 | 2,12 | 3,78 | 6,36 | 4,40 | 2,48 | 1,91 | 1,33 | 4,02 | 0,76 | 1,53 | 0,57 | 1,74 | 0,61 |

| Computing compromise ("Bod" vs. all) ranking list (Vector **CRL**) as **Lead** ×‖**PR**‖ | | | | | | | | | | | | | | | | | | | | | | | | | | | | | |
|---|---|---|---|---|---|---|---|---|---|---|---|---|---|---|---|---|---|---|---|---|---|---|---|---|---|---|---|---|---|
| Age | Avr | Bil | Bod | Cha | Chu | Dob | Dor | Ere | Evl | Fil | Gol | Gre | Gry | Hak | Kar | KoA | Kob | Las | Lem | Mas | Ner | Nev | Pol | Rub | She | Sok | Sto | TkV | Vin |
| 3,30 | 10,80 | 2,64 | 10,69 | 0,90 | 3,00 | 0,28 | 1,50 | 4,15 | 1,50 | 3,66 | 2,18 | 0,95 | 1,39 | 7,05 | 3,75 | 5,01 | 2,84 | 5,75 | 6,93 | 2,83 | 3,90 | 2,29 | 2,10 | 3,68 | 1,20 | 2,40 | 0,90 | 2,24 | 0,22 |

| Ordered compromise ("Bod" vs. all) ranking list (ordered Vector **CRL**) | | | | | | | | | | | | | | | | | | | | | | | | | | | | | |
|---|---|---|---|---|---|---|---|---|---|---|---|---|---|---|---|---|---|---|---|---|---|---|---|---|---|---|---|---|---|
| Avr | Bod | Hak | Lem | Las | KoA | Ere | Ner | Kar | Rub | Fil | Age | Chu | Kob | Mas | Bil | Sok | Nev | TkV | Gol | Pol | Dor | Evl | Gry | She | Gre | Cha | Sto | Dob | Vin |
| 10,80 | 10,69 | 7,05 | 6,93 | 5,75 | 5,01 | 4,15 | 3,90 | 3,75 | 3,68 | 3,66 | 3,30 | 3,00 | 2,84 | 2,83 | 2,64 | 2,40 | 2,29 | 2,24 | 2,18 | 2,10 | 1,50 | 1,50 | 1,39 | 1,20 | 0,95 | 0,90 | 0,90 | 0,28 | 0,22 |
| 1 | 2 | 3 | 4 | 5 | 6 | 7 | 8 | 9 | 10 | 11 | 12 | 13 | 14 | 15 | 16 | 17 | 18 | 19 | 20 | 21 | 22 | 23 | 24 | 25 | 26 | 27 | 28 | 29 | 30 |

**Figure 16.** Ranking list as a self-assessment compromise weighted by the leader = "Bod" (leader-staff compromise)

**Leader = "Avr"**

| "Avr" assesses ("gives weights") to the "experts" (vector **Lead**, which is the row "Avr" in the Matrix ‖**PR**‖) | | | | | | | | | | | | | | | | | | | | | | | | | | | | | |
|---|---|---|---|---|---|---|---|---|---|---|---|---|---|---|---|---|---|---|---|---|---|---|---|---|---|---|---|---|---|
| Age | Avr | Bil | Bod | Cha | Chu | Dob | Dor | Ere | Evl | Fil | Gol | Gre | Gry | Hak | Kar | KoA | Kob | Las | Lem | Mas | Ner | Nev | Pol | Rub | She | Sok | Sto | TkV | Vin |
| 3,21 | 10,99 | 2,63 | 10,86 | 0,88 | 2,92 | 0,29 | 1,46 | 4,09 | 1,56 | 3,60 | 2,14 | 0,93 | 1,38 | 7,31 | 3,84 | 4,93 | 2,78 | 5,61 | 6,89 | 2,87 | 3,80 | 2,34 | 2,05 | 3,85 | 1,17 | 2,34 | 0,88 | 2,19 | 0,22 |

| Computing compromise ("Avr" vs. all) ranking list (Vector **CRL**) as **Lead** ×‖**PR**‖ | | | | | | | | | | | | | | | | | | | | | | | | | | | | | |
|---|---|---|---|---|---|---|---|---|---|---|---|---|---|---|---|---|---|---|---|---|---|---|---|---|---|---|---|---|---|
| Age | Avr | Bil | Bod | Cha | Chu | Dob | Dor | Ere | Evl | Fil | Gol | Gre | Gry | Hak | Kar | KoA | Kob | Las | Lem | Mas | Ner | Nev | Pol | Rub | She | Sok | Sto | TkV | Vin |
| 3,36 | 10,36 | 2,68 | 10,68 | 0,92 | 3,05 | 0,25 | 1,53 | 4,12 | 1,51 | 3,64 | 2,17 | 0,96 | 1,35 | 7,12 | 3,73 | 5,00 | 2,87 | 5,85 | 6,96 | 2,74 | 3,97 | 2,31 | 2,14 | 3,69 | 1,22 | 2,44 | 0,92 | 2,26 | 0,19 |

| Ordered compromise ("Avr" vs. all) ranking list (ordered Vector **CRL**) | | | | | | | | | | | | | | | | | | | | | | | | | | | | | |
|---|---|---|---|---|---|---|---|---|---|---|---|---|---|---|---|---|---|---|---|---|---|---|---|---|---|---|---|---|---|
| Bod | Avr | Hak | Lem | Las | KoA | Ere | Ner | Kar | Rub | Fil | Age | Chu | Kob | Mas | Bil | Sok | Nev | TkV | Gol | Pol | Dor | Evl | Gry | She | Gre | Cha | Sto | Dob | Vin |
| 10,68 | 10,36 | 7,12 | 6,96 | 5,85 | 5,00 | 4,12 | 3,97 | 3,73 | 3,69 | 3,64 | 3,36 | 3,05 | 2,87 | 2,74 | 2,68 | 2,44 | 2,31 | 2,26 | 2,17 | 2,14 | 1,53 | 1,51 | 1,35 | 1,22 | 0,96 | 0,92 | 0,92 | 0,25 | 0,19 |
| 1 | 2 | 3 | 4 | 5 | 6 | 7 | 8 | 9 | 10 | 11 | 12 | 13 | 14 | 15 | 16 | 17 | 18 | 19 | 20 | 21 | 22 | 23 | 24 | 25 | 26 | 27 | 28 | 29 | 30 |

**Figure 17.** Ranking list as a self-assessment compromise weighted by the leader = "Avr" (leader-staff compromise)

However, the results of the third ranking option (according to the leader "Chu") debunk this assumption (Figure 18). The value system of "Chu" can be considered an alternative—the achievements of the category "Papers" (the most important according to the premium leaders) is rated as zero by "Chu."

Observably, all the three illustrative rankings differ little from each other. In the best case, we are talking about several castling on adjacent positions. It turns out that the figure of the leader plays an insignificant role in forming the compromise ranking. To verify this hypothesis, the experiment has been repeated for each employee in the role of an expert. These experiments showed similar results—the rankings constructed were almost identical.

**Leader = "Chu"**

"Chu" assesses ("gives weights") to the "experts" (vector **Lead**, which is the row "Chu" in the Matrix ‖**PR**‖)

| Age | Avr | Bil | Bod | Cha | Chu | Dob | Dor | Ere | Evl | Fil | Gol | Gre | Gry | Hak | Kar | KoA | Kob | Las | Lem | Mas | Ner | Nev | Pol | Rub | She | Sok | Sto | TkV | Vin |
|---|---|---|---|---|---|---|---|---|---|---|---|---|---|---|---|---|---|---|---|---|---|---|---|---|---|---|---|---|---|
| 4,12 | 6,53 | 3,05 | 9,85 | 1,12 | 3,75 | 0,00 | 1,87 | 4,12 | 1,28 | 3,75 | 2,25 | 1,12 | 1,12 | 6,80 | 3,21 | 5,25 | 3,37 | 7,12 | 7,28 | 1,93 | 4,87 | 2,36 | 2,62 | 3,00 | 1,50 | 3,00 | 1,12 | 2,62 | 0,00 |

Computing compromise ("Chu" vs. all) ranking list (Vector **CRL**) as **Lead** × ‖**PR**‖

| Age | Avr | Bil | Bod | Cha | Chu | Dob | Dor | Ere | Evl | Fil | Gol | Gre | Gry | Hak | Kar | KoA | Kob | Las | Lem | Mas | Ner | Nev | Pol | Rub | She | Sok | Sto | TkV | Vin |
|---|---|---|---|---|---|---|---|---|---|---|---|---|---|---|---|---|---|---|---|---|---|---|---|---|---|---|---|---|---|
| 3,41 | 10,02 | 2,72 | 10,66 | 0,93 | 3,10 | 0,23 | 1,55 | 4,10 | 1,51 | 3,63 | 2,16 | 0,97 | 1,33 | 7,17 | 3,71 | 5,00 | 2,90 | 5,93 | 6,98 | 2,67 | 4,03 | 2,33 | 2,17 | 3,69 | 1,24 | 2,48 | 0,93 | 2,28 | 0,18 |

Ordered compromise ("Chu" vs. all) ranking list (ordered Vector **CRL**)

| Bod | Avr | Hak | Lem | Las | KoA | Ere | Ner | Kar | Rub | Fil | Age | Chu | Kob | Bil | Mas | Sok | Nev | TkV | Pol | Gol | Dor | Evl | Gry | She | Gre | Cha | Sto | Dob | Vin |
|---|---|---|---|---|---|---|---|---|---|---|---|---|---|---|---|---|---|---|---|---|---|---|---|---|---|---|---|---|---|
| 10,66 | 10,02 | 7,17 | 6,98 | 5,93 | 5,00 | 4,10 | 4,03 | 3,71 | 3,69 | 3,63 | 3,41 | 3,10 | 2,90 | 2,72 | 2,67 | 2,48 | 2,33 | 2,28 | 2,17 | 2,16 | 1,55 | 1,51 | 1,33 | 1,24 | 0,97 | 0,93 | 0,93 | 0,23 | 0,18 |
| 1 | 2 | 3 | 4 | 5 | 6 | 7 | 8 | 9 | 10 | 11 | 12 | 13 | 14 | 15 | 16 | 17 | 18 | 19 | 20 | 21 | 22 | 23 | 24 | 25 | 26 | 27 | 28 | 29 | 30 |

**Figure 18.** Ranking list as a self-assessment compromise weighted by the leader = "Chu" (leader-staff compromise)

The only thing that distinguished them was the place in the ranking where the disturbances occurred (Figure 19). If the premium leader determines the value system ("Bod" or "Avr"), then small castling is observed at the top positions of the ranking. If the value system is determined in the opinion of the middle peasant (like "Chu"), fluctuations occur in positions in the second half of the ranking list.

**Ranking position**

| 1 | 2 | 3 | 4 | 5 | 6 | 7 | 8 | 9 | 10 | 11 | 12 | 13 | 14 | 15 | 16 | 17 | 18 | 19 | 20 | 21 | 22 | 23 | 24 | 25 | 26 | 27 | 28 | 29 | 30 |
|---|---|---|---|---|---|---|---|---|---|---|---|---|---|---|---|---|---|---|---|---|---|---|---|---|---|---|---|---|---|

Ordered Vector **CRL**("Bod")

| Avr | Bod | Hak | Lem | Las | KoA | Ere | Ner | Kar | Rub | Fil | Age | Chu | Kob | Mas | Bil | Sok | Nev | TkV | Gol | Pol | Dor | Evl | Gry | She | Gre | Cha | Sto | Dob | Vin |
|---|---|---|---|---|---|---|---|---|---|---|---|---|---|---|---|---|---|---|---|---|---|---|---|---|---|---|---|---|---|
| 10,802 | 10,695 | 7,051 | 6,925 | 5,753 | 5,008 | 4,146 | 3,898 | 3,749 | 3,679 | 3,658 | 3,298 | 2,998 | 2,840 | 2,826 | 2,643 | 2,399 | 2,287 | 2,240 | 2,176 | 2,099 | 1,499 | 1,497 | 1,389 | 1,199 | 0,947 | 0,900 | 0,900 | 0,283 | 0,217 |

Ordered Vector **CRL**("Avr")

| Bod | Avr | Hak | Lem | Las | KoA | Ere | Ner | Kar | Rub | Fil | Age | Chu | Kob | Mas | Bil | Sok | Nev | TkV | Gol | Pol | Dor | Evl | Gry | She | Gre | Cha | Sto | Dob | Vin |
|---|---|---|---|---|---|---|---|---|---|---|---|---|---|---|---|---|---|---|---|---|---|---|---|---|---|---|---|---|---|
| 10,679 | 10,363 | 7,122 | 6,957 | 5,852 | 4,999 | 4,117 | 3,969 | 3,731 | 3,689 | 3,643 | 3,358 | 3,053 | 2,874 | 2,737 | 2,684 | 2,443 | 2,315 | 2,264 | 2,169 | 2,137 | 1,527 | 1,508 | 1,354 | 1,221 | 0,958 | 0,916 | 0,916 | 0,253 | 0,194 |

Ordered Vector **CRL**("Chu")

| Bod | Avr | Hak | Lem | Las | KoA | Ere | Ner | Kar | Rub | Fil | Age | Chu | Kob | Bil | Mas | Sok | Nev | TkV | Pol | Gol | Dor | Evl | Gry | She | Gre | Cha | Sto | Dob | Vin |
|---|---|---|---|---|---|---|---|---|---|---|---|---|---|---|---|---|---|---|---|---|---|---|---|---|---|---|---|---|---|
| 10,659 | 10,020 | 7,167 | 6,983 | 5,932 | 4,995 | 4,097 | 4,027 | 3,713 | 3,689 | 3,634 | 3,408 | 3,098 | 2,903 | 2,716 | 2,667 | 2,478 | 2,334 | 2,283 | 2,169 | 2,165 | 1,549 | 1,513 | 1,327 | 1,239 | 0,968 | 0,929 | 0,929 | 0,230 | 0,176 |

**Figure 19.** Small fluctuations in the ranking lists with different choice of a leader

### 4.3. Comparing and mixing the rankings produced by different algorithms

We define the place-difference ($PLACE\_DIFF$) between two ordered ranking lists $\text{RankingList}_i$ and $\text{RankingList}_j$ constructed on top of the group of $m$ staff members as the following sum:

$$PLACE_{DIFF}(\text{RankingList}_i, \text{RankingList}_j) = \\ = \sum_{k=1}^{m} \left| PLACE_{\text{RankingList}_i}(k) - PLACE_{\text{RankingList}_j}(k) \right|, \qquad (9)$$

where $PLACE_{\text{RankingList}_i}(k)$, and $PLACE_{\text{RankingList}_j}(k)$ are the positions (ordinal numbers) of the $k$-th person in the $i$-th and $j$-th ranking lists, respectively.

The maximal possible place-difference ($\max_m PLACE\_DIFF$) between two ordered ranking lists of the group of $m$ staff members is the number which indicates the difference when the two compared lists are in the reverse order to each other:

$$\max_m PLACE\_DIFF = \left\lfloor \frac{m^2}{2} \right\rfloor. \qquad (10)$$

Finally, the place-distance ($PLACE\_DISTANCE$) between two ordered ranking lists of the group of $m$ staff members $\text{RankingList}_i$ and $\text{RankingList}_j$ can be defined as follows:

$$PLACE\_DISTANCE(\text{RankingList}_i, \text{RankingList}_j) = \frac{PLACE\_DIFF(\text{RankingList}_i, \text{RankingList}_j)}{\max_m PLACE\_DIFF}. \qquad (11)$$

For example, in Figure 11, we compared three ranking lists: administrative, democratic self-assessment, and "Social lift." Numeric pairwise comparison using formula (9) gives us the following three place-differences:

$PLACE\_DIFF(\text{AdministrativeRankingList}, \text{SelfAssessmentRankingList}) = 16;$

$PLACE\_DIFF(\text{AdministrativeRankingList}, \text{SocialLiftRankingList}) = 52;$

$PLACE\_DIFF(\text{SelfAssessmentRankingList}, \text{SocialLiftRankingList}) = 50.$

Comparison shown in Figure 15 adds yet another difference as follows:

$PLACE\_DIFF(\text{AdministrationStaffCompromiseRankingList}, \text{AdministrativeRankingList}) = 32.$

The maximal possible difference is $\max_{30} PLACE\_DIFF = 450$, therefore,

$PLACE\_DISTANCE(\text{AdministrativeRankingList}, \text{SelfAssessmentRankingList}) = 0.03(5);$

$PLACE\_DISTANCE(\text{AdministrativeRankingList}, \text{SocialLiftRankingList}) = 0.11 (5);$

$PLACE\_DISTANCE(\text{SelfAssessmentRankingList}, \text{SocialLiftRankingList}) = 0.(1);$

$PLACE\_DISTANCE(\text{AdministrationStaffCompromiseRankingList}, \text{AdministrativeRankingList}) = 0.07(1).$

The purely administrative and the purely personnel ranking lists differ with just 16 points (place-distance ≈ 3.6%); however their compromises (administration vs. personnel) differ from the administrative ranking list with more points (32, i.e., place-distance ≈ 7.1%). This means that, in this compromise, the voice of experts (those who get more weight) plays the decisive role and impacts the results.

We can also define more precise metric to measure differences between the ranking lists, which will use also concrete scores of the personnel and not only the ordinal numbers in the ranking lists. Let us define the score-difference ($SCORE\_DIFF$) between two ordered and normalized ranking lists, $\text{RankingList}_i$ and $\text{RankingList}_j$, constructed on top of the group of $m$ staff members as the following sum:

$$SCORE\_DIFF(\text{RankingList}_i, \text{RankingList}_j) = \sum_{k=1}^{m} \left| SCORE_{\text{RankingList}_i}(k) - SCORE_{\text{RankingList}_j}(k) \right|, \quad (12)$$

where $SCORE_{\text{RankingList}_i}(k)$ and $SCORE_{\text{RankingList}_j}(k)$ are the ranking scores (rational numbers between 0 and 1) of the $k$-th person in the $i$-th and $j$-th ranking lists, respectively.

The maximal possible score-difference ($\max_m SCORE\_DIFF$) between two normalized ranking lists of the group of $m$ staff members is the number that is equal to 2, i.e., $\max_m SCORE\_DIFF = 2$.

Finally, the score-distance ($SCORE\_DISTANCE$) between two ordered ranking lists of the group of $m$ staff members, $\text{RankingList}_i$ and $\text{RankingList}_j$, can be defined as follows:

$$SCORE\_DISTANCE(\text{RankingList}_i, \text{RankingList}_j) = 0.5 \cdot SCORE\_DIFF(\text{RankingList}_i, \text{RankingList}_j). \quad (13)$$

The use case and the example for the score-distances will be shown in the next Section.

We suggest also another way to compute the compromises in the personnel ranking based on iteratively mixing different algorithms during the ranking process. We name such procedures as *Compromise Dichotomy*. The compromise dichotomy ranking process is arranged as follows. We chose two ranking algorithms ($A$ and $B$) from the list already considered above. Then we made first ranking iteration using algorithm $A$, obtained ordered ranking list, and divided it into two equal parts (WINNERS and LOSERS). If number of ranked personnel is odd, then the WINNERS part will be one person longer than the LOSERS part. After that, we applied ranking algorithm $B$ independently for each of the two subgroups (WINNERS and LOSERS) and divided each subgroup to WINNERS and LOSERS the same way as above. We obtained four subgroups to which we again applied the algorithm, and so on, until the number of persons within the subgroups became equal to one. After that, we concatenated the subgroups to the resulting compromise ranking list. The number of iterations for the initial group of $m$ persons would be $\lceil \log_2 m \rceil$.

One may see that the choice of the algorithm for the first (and all odd ones) iteration makes this algorithm dominant in the compromise. For example, if a person from the group of 64 persons gets ranking score from $A$ algorithm, which puts the person to the 33$^{rd}$ place (i.e., to the LOSERS part), then this person will not get better than 33$^{rd}$ place after all of the following iterations.

The most interesting pair of the algorithms for the compromise dichotomy is the Administrative Ranking vs. Staff Self-Ranking and, depending on the choice of the algorithm for the first (odd) iteration, we have the two following options for the dichotomy process:

- *Weak Compromise Dichotomy*—all the odd iterations are made using the administrative value system for the ranking, and all the even ones are performed as personnel self-assessment (see Figure 20). The resulting ranking list would be $WCDR$ (Weak, Compromise, Dichotomy, Ranking);

- *Strong Compromise Dichotomy*—this is a more democratic compromise, where all the odd iterations are ranked based on group(s) self-assessment and all the even iterations used administrative value system for ranking. The resulting ranking list would be $SCDR$ (Strong, Compromise, Dichotomy, Ranking).

There is also one possible way to use smartly the dichotomy process, which is a *Self-Compromise Dichotomy*. Such process uses just one ranking algorithm (the best choice is group self-assessment) and applies it iteratively within the subgroups the same way as explained above. This is a very interesting method of complete democracy, which tries to make the result more accurate (free from the noise of losers) at each next iteration. The resulting ranking list would be $CDSR$ (Compromise, Dichotomy, Self-Ranking). Many administrative decisions in NURE have been made based on such transparent democracy as the Self-Compromise Dichotomy ranking.

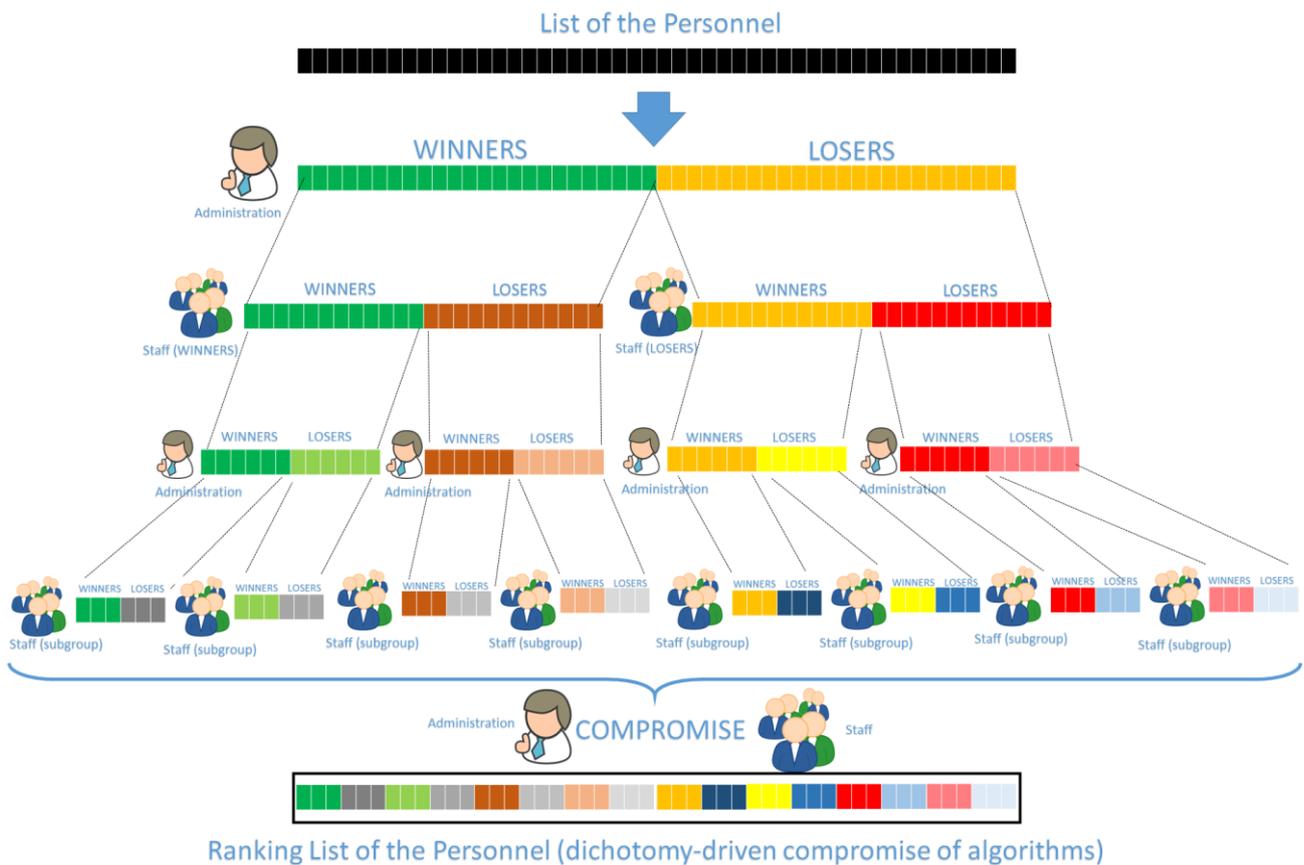

**Figure 20.** Four iterations of the Week-Compromise Dichotomy ranking process illustrated (odd iterations—assessment by the administration; even iterations—the group self-assessment)

To weaken the effect of the first iteration, one can apply also the *League-Driven Dichotomy* ranking process when there is a possibility to exchange (after each iteration) the last loser(s) from former winners with the first winner(s) from the former losers leaving the chance for recovery for someone after the unfortunate first iteration.

Finally, it is good to notice that other possible options for all these compromises would be using the more tolerant *Golden Ratio* instead of *Dichotomy* division principle when (at each iteration) the WINNERS subgroups exceed the LOSERS subgroups.

## 5. Rewards and Assessment of Reinforcements

Assume that we have some (academic) organization, which personnel comprising $m$ (academic) staff members (*object of assessment*). In our further experiments, we used data from NURE.

Assume that each staff member is assessed based on his/her rewards (bonuses, incentives, encouragements, promotions, honors, powers, titles, additional resources, improved working and wellbeing conditions, awards, and other reinforcements) obtained from the University (or related to the work for the University) during a certain period. All the accountable achievements are divided into $k$ predefined categories. The aggregated reward obtained by each staff member is the number of documented (registered) rewards of each $k$ category.

All the reinforcements' statistics regarding all the staff members of the organization are represented by matrix $Bp$ (Bonuses, personnel) as $[m \times k]$. For example, the value $Bp(i,j) = 3$, would mean that the evidence on 3 bonuses of category $j$ obtained by $i$-th staff member exist, according to the registered records.

### 5.1. Reward Value Systems and Analytics of Reinforcements

Assume that the administration of the organization (*subject of assessment*) publishes its "*value system*" as an instrument of the transparent reinforcement of the academic personnel. The administrative value system for the reinforcements is a vector $RVSa$ (<u>R</u>einforcement <u>V</u>alue <u>S</u>ystem, <u>a</u>dministration) or a row $[1 \times k]$, which indicates the relative value the administration gives for each instance of the reward of a certain category. The vector is normalized as follows: $\sum_{i=1}^{k} RVSa(i) = 1$. For example, $RVSa(2) = 0.2$ would mean that the administration considers the relative value of the 2$^{nd}$ category of rewards as 20% of the overall value.

$ABR$ (<u>A</u>dministration, <u>B</u>onuses, <u>R</u>anking) is a vector (a one-row matrix $[1 \times m]$), which indicates the reinforcements' assessment score of each $m$ staff member following the administrative value system. This vector is computed (using similar schema as the one defined in Formula (1)) as the multiplication of the administrative reinforcement value system vector matrix $[1 \times k]$ with the transpose matrix $[k \times m]$ of the actual rewards obtained by the staff members as follows:

$$ABR = RVSa \times Bp^T. \qquad (14)$$

It is evident that the reinforcements will work if their value will be adequately perceived and appreciated by the staff, to whom these rewards are being applied, and each of whom may have his/her opinion on the value of each reward for oneself. Therefore, a *self-assessment* regarding own rewards is even more important for each staff member than self-assessment of own achievements. Each staff member of the organization publishes its personal "*value system*" as a transparent instrument for the self-assessment regarding the bonuses obtained, and all such individual value systems are collected together as a matrix. The value system of the personnel regarding the reinforcements is a matrix $RVSp$ (<u>R</u>einforcement <u>V</u>alue <u>S</u>ystem, <u>p</u>ersonnel) as $[m \times k]$, which indicates the value of relative importance that a particular staff member (the matrix row) gives for each reward of a certain category (the matrix column). The vector is normalized as follows: $\forall j (j = \overline{1,m}), \sum_{i=1}^{k} RVSp(j,i) = 1$. For example, $RVSp(4,3) = 0.18$ would mean that person number 4 considers the relative importance of the 3$^{rd}$ category of the awards as 18% of the overall importance.

$PBR$ (Personnel, Bonuses, Self-Ranking) is a matrix $[m \times m]$, which indicates the reward-related assessment score of each $m$ staff member following the personal reinforcement value system of each $m$ staff member. This matrix is computed (similar as it is done in Formula (2)) as the multiplication of the personnel reinforcement value system matrix $[m \times k]$ with the transpose matrix $[k \times m]$ of the actual rewards obtained by the staff members as follows:

$$PBR = RVSp \times Bp^T. \qquad (15)$$

One can similarly obtain the normalized rankings as we did before, i.e., $ABA = \|ABR\|$, using formula (4), $\|PBR\|$ using Formula (5), and $DBA$ using formula (6), where $ABA$ (Administrative Bonuses Assessment) would be the ranking list of the personnel regarding the reinforcements constructed by the administration (similar to $AA$); and $DBA$ (Democratic Bonuses Assessment) would be the ranking list of the personnel regarding the reinforcements constructed by the democratic procedure of self-assessment (similar to $DA$).

One may easily notice a certain analogy (like a mirror reflection): (a) the matrix $Rp$, which indicates the distribution of achievements among the academic staff, corresponds to the matrix $Bp$, which indicates the distribution of bonuses among the academic staff; (b) the vector $AVSa$, which is an administrative value system to assess the achievements, corresponds to the vector $RVSa$, which is an administrative value system to assess the bonuses (reinforcements); (c) the matrix $AVSp$, which is a collection of personal value systems used for self-assessments of achievements, corresponds to the vector $RVSp$, which is a collection of personal value systems used for self-assessments of the bonuses (reinforcements). Due to this analogy, we can similarly easily define new (reward-based) ranking lists as we define such lists based on achievements. Table 1 contains the correspondence between the eight previously defined (achievement-based) ranking lists and the eight "mirror" ranking lists (reward-based), which are reasonable to compute and apply the same way using the same algorithms (analytics), which were already presented in the paper.

**Table 1.** Correspondence between the ranking lists based on staff achievements and the ranking lists based on distribution of bonuses among staff

| Ranking Lists for the Personnel | Achievements-Based | | Rewards-Based |
|---|---|---|---|
| Administrative Ranking List | $AA$ | formulas (1) and (4) | $ABA$ |
| Democratic Self-Assessment Ranking List | $DA$ | formulas (2), (5) and (6) | $DBA$ |
| Social Lift Ranking List | $SLR$ | Case 2 | $SLBR$ |
| Administration-Staff Compromise Ranking List | $CR$ | formula (7) | $CBR$ |
| Leaders-Staff Compromise Ranking list | $CRL$ | formula (8) | $CRBL$ |
| Weak Compromise Dichotomy Ranking List | $WCDR$ | paragraph 4.3, Figure 20 | $WCDBR$ |
| Strong Compromise Dichotomy Ranking List | $SCDR$ | paragraph 4.3 | $SCDBR$ |
| Self-Compromise Dichotomy Ranking List | $CDSR$ | paragraph 4.3 | $CDSBR$ |

## 5.2. Collision of Ranking Lists and Analytics of Justice

Organizational (procedural) justice, as the extent to which leaders apply fair procedural rules in the decision-making process, is highly important for both employees and organizations (Zheng et al., 2020). For a healthy climate within the organization, it is important that distribution of rewards among the staff members well corresponds to the distribution of their achievements. Therefore, study of collisions between the ranking lists from the "achievement-based" group and the "reward-based" group (presented in Table 1) would be the best way to assess the level of "justice" or "injustice" within the organization.

We define the measure of injustice within the organization $INJUSTICE(ARank, BRank)$ based on collision of two ranking lists ($ARank$ belongs to the "achievement-based" group, $BRank$—to the "reward-based" group) as the following rational number between 0 (complete justice) and 1 (complete injustice) as follows:

$$INJUSTICE(ARank, BRank) = SCORE\_DISTANCE(ARank, BRank), \quad (16)$$

where $SCORE\_DISTANCE$ is defined in Formula (13).

One may notice that the number of different measurements for the justice using all the ranking lists from Table 1 would be $8 \times 8 = 64$.

We also define a measure for the overall injustice $OVERALL\_INJUSTICE$, or the number, which aggregates the injustice measures (Formula (16)) for all possible pairs of available ranking lists as their contra-harmonic mean:

$$OVERALL\_INJUSTICE = \frac{\sum_{\forall i,j}[INJUSTICE(ARank_i, BRank_j)]^2}{\sum_{\forall i,j} INJUSTICE(ARank_i, BRank_j)}. \quad (17)$$

Both measurements (Formula (16) and Formula (17)) can be represented in percent (%) simply multiplying by 100.

### Description of the experiment (usage example)

Let us provide an example with the same organization we considered before in this paper. We will limit the example with the two ranking lists from the "achievement-based" group ($AA$ and $DA$) and two ranking lists from the "reward-based" group ($ABA$ and $DBA$). Therefore, we will compute four different measurements for the injustice within the organization and one overall injustice measure.

We have the following inputs for the experiment:

- Ranking lists $AA$ and $DA$, which have been already computed earlier in this paper for the group of 30 staff members;
- Matrix $Bp$ (Bonuses, personnel), i.e., $[30 \times 3]$, which shows actual distribution of the bonuses/rewards (of 3 types: "Salary," "Advancements," and "Awards") among the 30 staff members (see Figure 21);
- Vector $RVSa$ (Reinforcement Value System, administration), i.e., $[1 \times 3]$, which shows the relative importance of different 3 categories of rewards according to the opinion of the administration (Figure 21);

- Matrix $RVSp$ (Reinforcement Value System, personnel), i.e., $[30 \times 3]$, which shows the relative importance of different 3 categories of rewards according to the opinion of each of 30 staff members (Figure 21).

Computed "injustice" evaluations (four basic ones: $INJUSTICE(AA, ABA)$; $INJUSTICE(DA, DBA)$; $INJUSTICE(AA, DBA)$; and $INJUSTICE(DA, ABA)$ and the overall one: $OVERALL\_INJUSTICE$) regarding the newly computed ranking lists $ABA$ and $DBA$ vs. previously computed ranking lists $AA$ and $DA$ are presented in Figure 22.

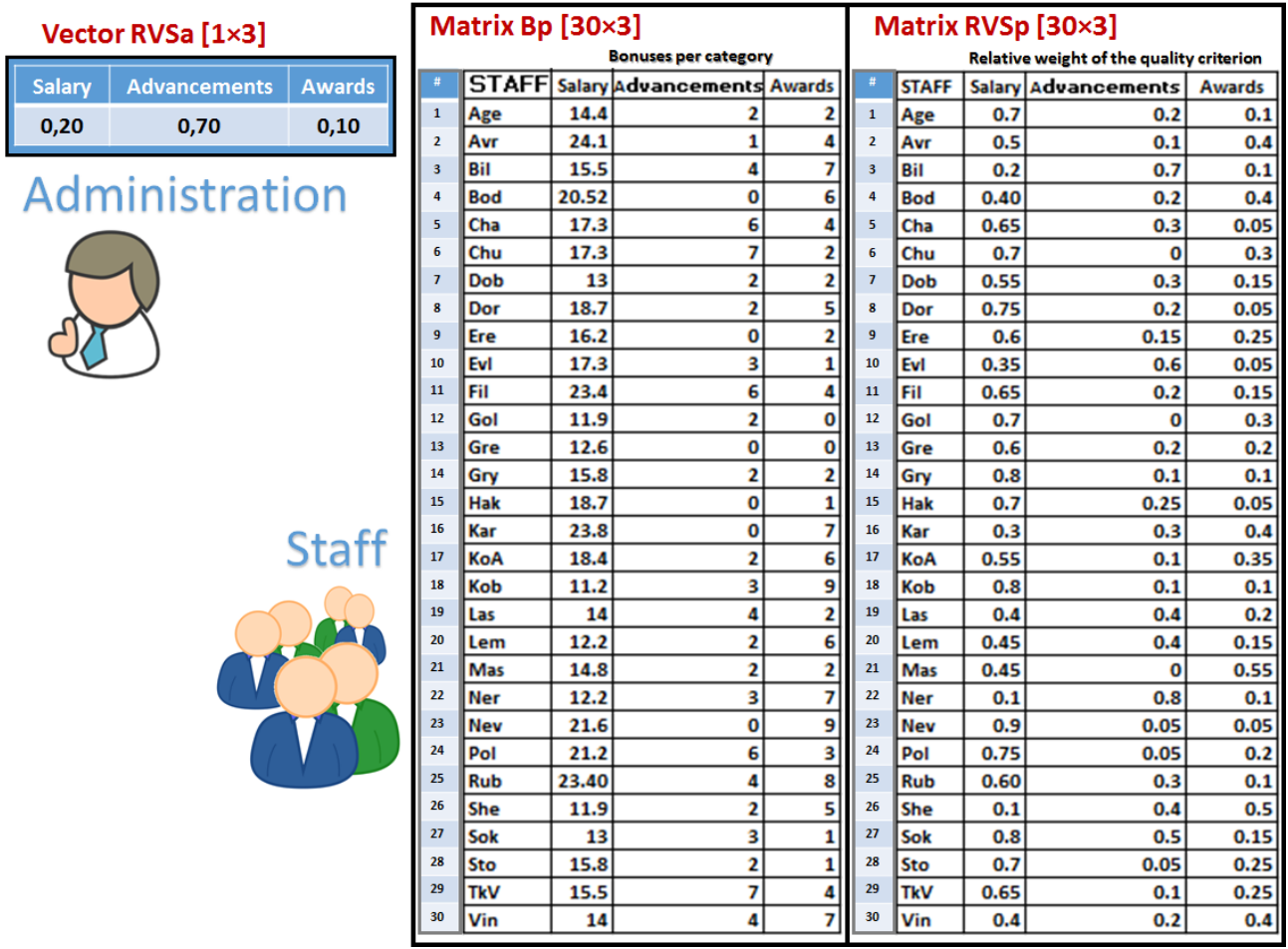

**Figure 21.** Input data for the experiments with reinforcements (reinforcement value systems *RVSa*—"administration" and *RVSp*—"personnel" and the reward (three categories) distribution among the staff *Bp*)

High value for the $INJUSTICE(AA, ABA)$ usually indicates that the administration does not reward for the achievements it formally requires from the personnel, i.e., it acts according to the destructive principle of "punishments for the innocent and rewards for the uninvolved." High value for the

$INJUSTICE(DA, DBA)$ usually indicates that the staff feel that their accomplishments do not meet the rewards they deserve. High value for the $INJUSTICE(AA, DBA)$ usually indicates that everyone is dissatisfied with what they receive: the administration is dissatisfied with the achievements, and the staff—with the bonuses. High value for the $INJUSTICE(DA, ABA)$ usually indicates that everyone overestimates their merits: the staff members think that they work very hard, and the administration thinks that it rewards very coolly (the principle "they pretend to pay, but we pretend to work"). Finally, the high value for the $OVERALL\_INJUSTICE$ usually indicates the overall poor emotional climate within the organization due to the lack of justice regarding the reinforcement instrument (reward allocation).

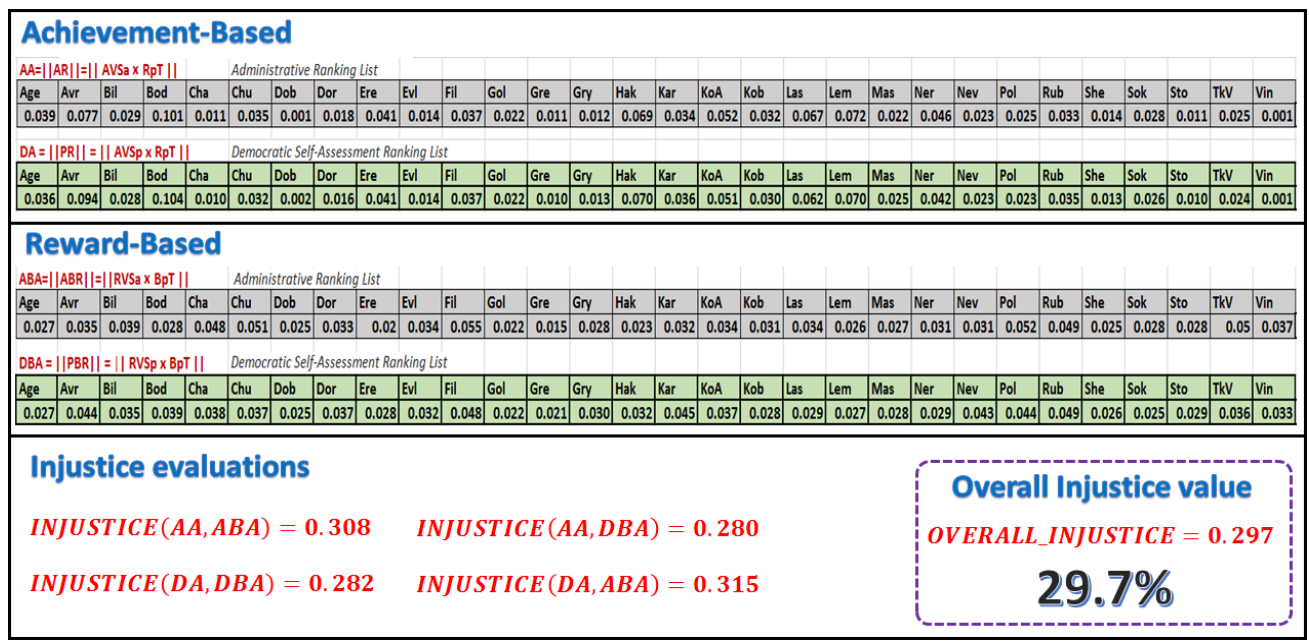

**Figure 22.** Computed achievement-based and reward-based ranking lists and the values of "injustice" regarding each pair of ranking lists and the "overall injustice" for the experiment

## 6. Work Passion (computing and ranking)

**Brief**

Another important measurement related to the key organizational behavior concepts would be the work passion. Passion can be defined as a strong feeling toward a personally important value/preference that motivates intentions and behaviors and produces beneficial effects on personal performance being "something" that drives us (Jachimowicz et al., 2018). Therefore, captured knowledge about work passion within an organization may play an important role in predicting the organizational performance and could also be used as an additional feature, which can contribute to the assessment and selection processes within the organization.

**Intuition behind the approach**

Work passion in academic organization can be seen as an intense feeling of an employee toward achieving something personally valuable at work, which then pushes his/her toward a certain behavior (creative work on academic objectives) within the organization. Therefore, one way to measure this for a particular person and to rank the personnel accordingly would be finding the correlation between personal values and actual achievements regarding the rewards. Most existing efforts on quantifying passion into some numeric scales are based on questionnaires (Sigmundsson et al., 2020); thus the passion assessment processes are essentially subjective. However, we are going to present an analytical (evidence-based, mechanical) procedure to measure the work passion. We suggest a computational approach to the work passion. It will be based on the basic organizational data (personal achievements, personal rewards, and personal values), and it will indicate for each person how big is the share of the intrinsic motivation of the person in the activity toward creating academic values comparably to the extrinsic motivation.

**Procedures**

We present the following self-assessment procedure to compute the work passion for each employee within the academic organization and to rank the whole academic personnel following their work passion.

Passion computation procedure (schema) is as follows:

$$WP = \|((Rp \odot AVSp) \times Rp^T) \oslash ((Bp \odot RVSp) \times Bp^T)\|, \tag{18}$$

where:

- $WP$ (Work, Passion) is a normalized (from 0 to 100%) square matrix $[m \times m]$, which contains the work passion score (share within the group) regarding each $m$ employee according to the value systems (for the achievements and for the rewards) of all the $m$ employees (including oneself). For example, the value $WP(\text{"Mary," "John"}) = 5$ within the group of $m = 30$ persons means that "$Mary$" would give to "$John$" the evaluation score for his work passion to be as much as 5% of the work passion of all the group of the employees, i.e., "$John,$" according to "$Mary,$" is more passionate than average person within the group ($5\% > 3.33\%$);
- ×—matrix product;
- ⊙—element-wise multiplication of matrices;
- ⊘—element-wise division of matrices.

**Comments on the procedure**

The stage (computing the matrix of convolved value systems regarding achievements), $S1 = Rp \odot AVSp$, can be seen as an evaluation of a mutual effect (aka mathematical convolution) of each personal value system (regarding types of achievements) to the corresponding person's performance (actual achievements) and vice versa.

The stage (computing the achievements' self-evaluation matrix based on convolved value systems regarding rewards), $S2 = S1 \times Rp^T$, gives the evaluations (each one by each one) of the fitness of the declared values to the actual achievements (important indicator of the work passion correlated with the intrinsic motivation of a person);

The stage (computing the matrix of convolved value systems), $S3 = Bp \odot RVSp$, can be seen as an evaluation of a mutual effect (also known as mathematical convolution) of each personal value system

(regarding types of rewards) to the corresponding person's encouragement (actual rewards) and vice versa;

The stage (computing the rewards' self-evaluation matrix based on convolved value systems), $S4 = S3 \times Bp^T$, gives the evaluations (each one by each one) of the fitness of the declared values to the actual rewards (important component of the work passion correlated with the extrinsic motivation of a person);

The stage (computing the ratio of the intrinsic vs. extrinsic motivation indicators), $S5 = S2 \oslash S4$, gives the evaluations (each one by each one) of the work passion.

The stage (normalization to 100%), $WP = \|S5\|$, gives the share of each individual work passion within the aggregated work passion of the whole group.

**Description of the experiment (usage example)**

Evaluation of the work passion is completely an evidence-based procedure, and it takes as an input the basis evidence on our target organization (we used to demonstrate our experiments):

- Matrix $Rp$ (Results, personnel), i.e., $[30 \times 4]$ matrix, contains the academic achievements' statistics (divided into 4 categories) from 30 staff members of the university (Figure 3);
- Matrix $AVSp$ (Academic Value System, personnel), i.e., $[30 \times 4]$ matrix, indicates the value of relative importance that a particular staff member (row of the matrix) gives for each academic achievement of a certain category (columns of the matrix correspond to 4 categories) (Figure 4b);
- Matrix $Bp$ (Bonuses, personnel) in our experiment is a $[30 \times 3]$ matrix, which shows evidence on actual rewards of different 3 categories obtained by each of 30 staff members (Figure 21);

- Matrix $RVSp$ (Reinforcement Value System, personnel), i.e., $[30 \times 3]$ matrix, shows the relative importance of different 3 categories of rewards following the opinion of each of 30 staff members (Figure 21).

Figure 23 shows the results of the experiment (work passion self-assessment matrix and the corresponding ranking list of the personnel).

**Figure 23.** Computed work passion self-evaluation matrix (each one evaluates each one, including oneself) and final work passion ranking list (ordered average values from the self-assessment matrix) for the experiment.

## Observations and conclusive remarks

One may see from the resulting ranking list quite high deviation of the passion values and that the overall passion of the leaders (about 25% of the personnel in our experiment) is more than the overall passion of the other 75% (rest of the group). It is an interesting message to the managers that actual drivers of potential success of the organization are concentrated in the hands and hearts of the quarter of the whole personnel. Combining such ranking list with other ranking lists (as shown in Subsection

4.3) may give quite valuable analytical evidence on correlation of the organizational outcomes (achievement scores) with the organizational values and management culture (rewards, justice, passion, etc.).

## 7. Conclusions

In this paper, we explore the concept of deep evidence and its impact on the decision-making processes (related to assessment and selection) within academic organizations (particularly within universities). We suggest ranking people not only based on their academic portfolios (i.e., evidence on academic achievements) but also using the evidence on their reinforcements (various rewards obtained from the organization). However, we believe that to make the evidence really "deep," we need to analytically contextualize it through various personal views (opinions, preferences). Therefore, we suggest also making explicit individual values (personal preferences as weights in the assessment formulas), which can be used as formal specifications of personal (self-)assessment procedures regarding achievements and rewards. Collected deep evidence (achievement portfolio, reward portfolio, and value portfolio for everyone within the organization) enables various assessments and selection procedures with high-level transparency and different balances of decision power, which we have suggested in this paper.

Throughout the paper, we provided our observations and conclusive remarks where we addressed our research questions separately for each suggested analytical procedure, i.e., arguing on the benefit of applying the deep evidence for the evidence-based assessment, ranking, and selection, along with measuring the democracy, justice, and work passion within the organization.

We now summarize the backbone philosophy behind our study. Assume a selection procedure to be a kind of "black box," which takes the available evidence as an input and computes the scores for each possible decision (selection) option as an output. If the people who are the object of selection do not know all the evidence, except their own, and they are only informed about the outcome, then the procedure cannot be considered transparent at all. Assume that both (all the evidence at the input and scores from the output) are made public for all the people involved in the selection process, this will make the procedure transparent but quite insignificantly. Assume that, in addition to publishing input and output evidence, we also open the well-formalized decision procedure (decision-making "formula") to the public so that everyone can compute and check their own and others' scores, then this will be much more advanced transparency and will induce a better climate within the organization. However, can we do even more than that? A university is a space for accelerated personal development where both students and academic personnel co-develop, co-evolve, and co-create within different collaborative processes. Inferably, the key for the personal development of an employee is the belief (or feeling) that his/her personal opinion (view) is considered when the important (for him/her) organizational decisions are being made. Therefore, a weak point that remains in the above logic of transparency is that personal involvement is missing. Here, the deep evidence can help. We enable (encourage) everyone to publish (make available for the decision-making procedure as well for all the people involved) their own value systems (parameters or weights used in the decision "formula"). Hence, we are making an essential upgrade of the transparency because every decision maker and every member of the academic personnel openly and, in advance, show how he/she will value different decision-making criteria for a particular selection procedure. It is only when the (input) evidence is made available for all that the decision process will be not only transparent but also fair without possibilities for manipulations. Concurrently, different kinds of compromises will be possible in balancing among the value systems provided by the administration (or decision makers) and the personnel. Several options for such compromises are shown in this paper, and examples of their use and impact in the university are provided.

In this paper, we cover only a small part of the hidden potential of applying deep evidence in organizational processes. Our future plans include updates to the TRUST portal to enable autonomous digital (agent-driven) cognitive "clones" of the decision makers. Such "clones" will use the value systems of their "donors" and will be capable of 24/7 automated and transparent (deep) evidence-based decision-making at the portal. We plan to use our Pi-Mind cloning technology (Terziyan et al., 2018) to make a radical digital transformation of the assessment and selection processes within the universities, aiming to remove as much as possible available bias and corruption in the decision-making.